\documentclass[10pt, conference]{IEEEtran}
\usepackage[utf8]{inputenc} 

\usepackage{amsmath,amssymb,amsfonts}
\usepackage{graphicx}
\usepackage{textcomp}

\usepackage{amsthm}
\usepackage{xcolor}
\usepackage[figuresright]{rotating}

\usepackage{graphicx}
\graphicspath{{/}{fig/}}

\usepackage{array}
\usepackage{textcomp}
\usepackage{xcolor}
\usepackage{multirow}
\usepackage{booktabs}

\usepackage{mathtools}
\usepackage{breqn}
\usepackage{float}

\usepackage{pgfplots}
\pgfplotsset{compat=1.7}
\usepgfplotslibrary{groupplots}

\usepackage{multirow,tabularx}

\newlength\figureheight
\newlength\figurewidth
\title{\LARGE \bf
    Blockchain-Powered Collaboration in \\ Heterogeneous Swarms of Robots
}

\author{
    \IEEEauthorblockN{
        Jorge Peña Queralta and Tomi Westerlund
    }\\
    \IEEEauthorblockA{
        Turku Intelligent Embedded and Robotic Systems \\
        Faculty of Science and Engineering, University of Turku, Finland \\ 
        {\{jopequ, tovewe\}@utu.fi} \\
        {https://tiers.utu.fi}
    }
}

\begin{document}

\maketitle
\thispagestyle{empty}
\pagestyle{empty}

\begin{abstract}

    One of the key challenges in the collaboration within heterogeneous multi-robot systems is the optimization of the amount and type of data to be shared between robots with different sensing capabilities and computational resources. In this paper, we present a novel approach to managing collaboration terms in heterogeneous multi-robot systems with blockchain technology. Leveraging the extensive research of consensus algorithms in the blockchain domain, we exploit key technologies in this field to be integrated for consensus in robotic systems. We propose the utilization of proof of work systems to have an online estimation of the available computational resources at different robots. Furthermore, we define smart contracts that integrate information about the environment from different robots in order to evaluate and rank the quality and accuracy of each of the robots' sensor data. This means that the key parameters involved in heterogeneous robotic collaboration are integrated within the Blockchain and estimated at all robots equally without explicitly sharing information about the robots' hardware or sensors. Trustability is based on the verification of data samples that are submitted to the blockchain within each data exchange transaction, and validated by other robots operating in the same environment. Initial results are reported which show the viability of the concepts presented in this paper.

\end{abstract}


\IEEEpeerreviewmaketitle

\section{Introduction}

Autonomous robotic systems have significantly increased their penetration in multiple industries and research areas over the past decade, from healthcare \cite{van2016healthcare} to self-driving cars \cite{bojarski2016end}, including smart industry \cite{russmann2015industry}, or agriculture \cite{yaghoubi2013autonomous}. At the same time, part of the research focus has shifted from the design and development of complex robotic systems, such as humanoid robots \cite{kaneko2008humanoid}, to the design and development of methods and algorithms for cooperation in multi-robot systems \cite{alami1998cooperation, dadgar2016pso}. Through the interaction and collaboration in large robotic systems or swarms of robots \cite{gross2006autonomous}, simple individual operations can be combined towards swarm-level coordinated behavior and execution of higher-level tasks. In these systems, apparently simple or limited individual robots can exhibit complex behaviors at the swarm level \cite{rubenstein2012kilobot}.

\begin{figure*}
    \centering
    \includegraphics[width=.72\textwidth]{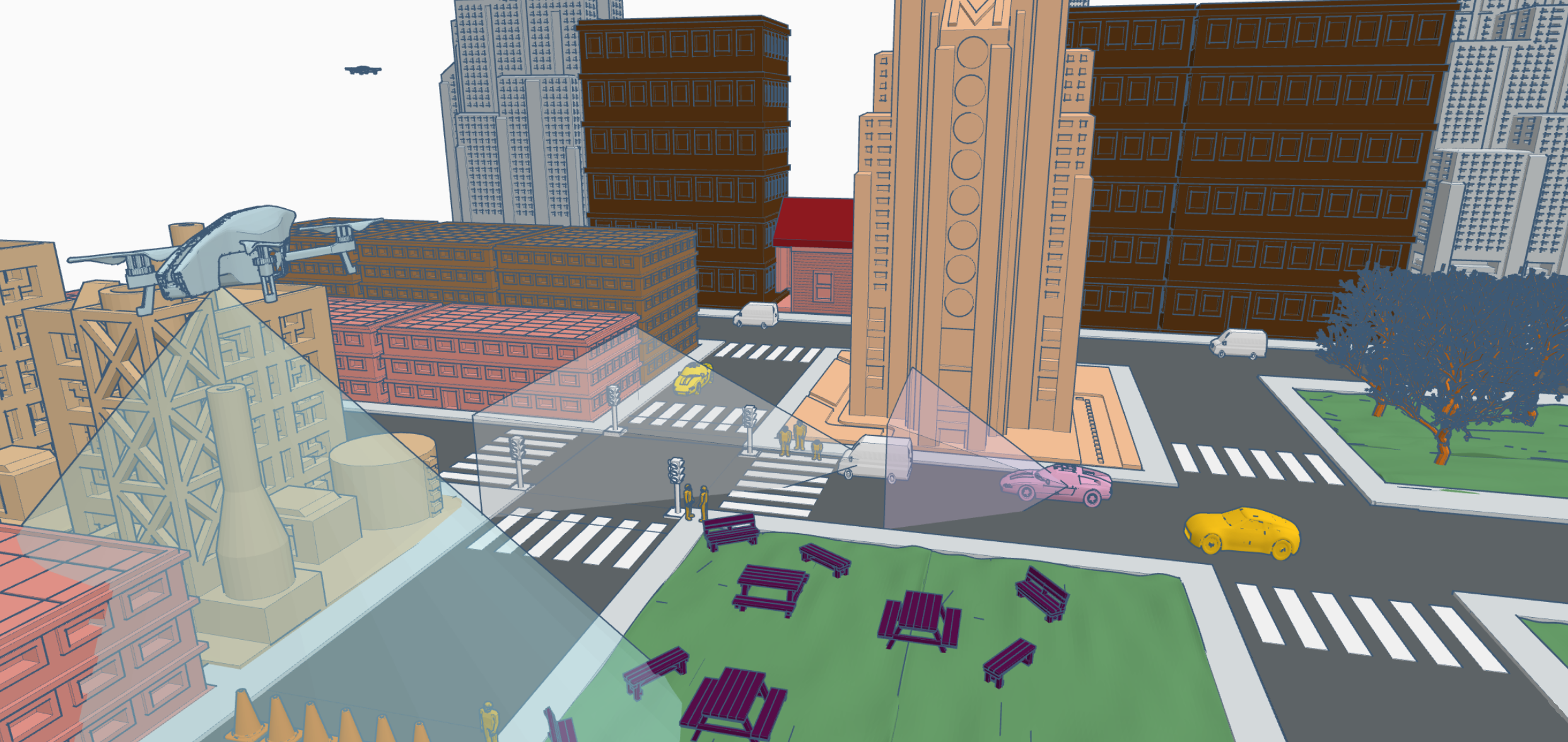}
    \caption{Illustration of collaborative sensing within a heterogeneous robotic system.}
    \label{fig:example}
\end{figure*}

Managing the collaboration in multi-robot systems presents multiple challenges. In the Kilobot project, \cite{rubenstein2012kilobot} utilize identical robots that can produce arbitrary two-dimensional geometric shapes through collaborative self-assembly. Therefore, the algorithms that run on the robots can be identical as well, and robots collaborate with any other robot in a predefined and, to some extent, static manner. However, in heterogeneous multi-robot systems, algorithms that require system-level consensus such as task allocation need to take into account the variable capabilities and resources of the different robots. \cite{shi2010task} introduced a reputation-based method to perform distributed task allocation, which significantly increases the systems' robustness, reliability and performance. Having a consensus method that effectively integrates the miscellaneous properties of operational characteristics, computational resources and sensing capabilities of heterogeneous robots in a swarm is therefore essential for optimizing consensus and the overall system performance. Figure \ref{fig:example} illustrates multiple autonomous vehicles and robots operating in the same environment. Through multi-robot collaboration, the white van is able to share data with the pink and yellow car coming behind of the pedestrians that are starting to cross the street. At the same time, the drone can share its data with the other vehicles for them to adapt their path planning algorithms when processing the images and concluding that the road is closed. In order to optimize collaboration in this scenario, the streaming of the drone's images towards the self-driving cars should be prioritized if compared with sending dense 3D lidar point clouds to the drone, as cars probably have enough processing power to analyze the images in real-time and benefit the most out of it. The drone, however, with a resource constrained on-board computer might be unable to process lidar data. Instead, if we provide a way for the cars to estimate the drone's available computational resources, then the data can be either downsampled or, if possible, preprocessed before sharing it.

Consensus theory in multi-agent systems can be considered for managing the collaboration in multi-robot systems \cite{qin2016recent}. In particular, consensus algorithms and protocols in blockchain systems are able to provide not only a decentralized and scalable solution to multi-robot consensus, but also raise the level of security and trustability within the collaborative system \cite{sankar2017survey, wang2019survey}. In fact, the integration of blockchain technologies within swarm robotic systems has been a subject of study over the past few years with an increasing interest from the research community \cite{ferrer2018blockchain, yuan2018blockchain}. Not only does a blockchain provide solutions for managing byzantine robots as in the work from \cite{strobel2018managing}, but also is an effective solution to solve the consensus problem \cite{nguyen2019blockchain}. In addition, parts of the blockchain stack can be exploited to secure specific aspects of collaboration that require sharing sensitive information between robots, such as the integration of \cite{castello2019merkle} of Merkle Trees for secure information sharing and mission progress reporting. This is a solution that has enormous potential to be utilized in a variety of robotic cooperation domains, from cooperative mapping \cite{queralta2019collaborative}, to search and rescue missions \cite{cardona2019robot}.

We consider the integration of blockchain technology for managing collaboration in heterogeneous multi-robot systems in order to solve existing challenges deriving from the potentially variable (i) sensing capabilities and (ii) computational resources in different robots. Because of this heterogeneity, robots not only generate a different amount and type of data but are also able to process it at different rates. Therefore, it is essential to have an automated way for robots to reach agreements in terms of both the format and size of data that is transmitted between the different collaborating robots, and the accuracy or trustability and reliability of the provided data. In this paper, we propose the utilization of a Proof of Work (PoW) scheme to have an online estimation of the available computational resources at different robots. In addition, the Proof or Work is combined with an algorithm that classifies the data provided by the different robots in terms of accuracy or usability and reduces the probability of tampered data being shared by a malicious attacker. This is then utilized in order to manage the bandwidth utilization in the peer-to-peer network utilized for robot-to-robot communication.

Blockchain platforms can be classified into two main types --- permissionless and permissioned. These can also be denominated public (permissionless), and consortium or private blockchains (permissioned). In public blockchains, there is no authority and all nodes are equivalent. In consortium or private blockchains, there are trusted authorities or nodes in charge of validating transactions \cite{zheng2017overview}. One of the key challenges in the utilization of a blockchain for applications requiring real-time communication and data processing is the scalability, as indicated by~\cite{ferrer2018blockchain}. This issue affects specially to permissionless or open blockchains, where all nodes in the network need to arrive to a consensus in order to validate each transaction. Nevertheless, recent advances in blockchain technology show promising results and potential for scalable and low-latency blockchain networks. \cite{luu2016secure} presented Elastico, where sharding in a permissionless blockchains was explored. Sharding is a technique that allows for distributed consensus in a network where nodes are divided in subnetworks or committees. Rather than processing and confirming all transactions globally across the network (for example through a majority consensus), each committee is in charge of processing a disjoint set of transactions, also denominated shard. In Elastico, researchers demonstrated the first sharding protocol that is secure in the presence of byzantine adversaries. \cite{kokoris2018omniledger} introduced OmniLedger, a decentralized and secure ledger that scales linearly with the size of the network and supports transaction confirmation times of under two seconds, potentially being able to match credit card standards in terms of transaction confirmation response time with a large enough network, compared to an average transaction confirmation time (block validation) of around ten minutes in the case of Bitcoin. While Elastico scales almost linearly with the available computation power in the network, OmniLedger does so with the  number of active validators.

In this paper, we propose an algorithm for managing collaboration in heterogeneous multi-robot system consisting on the following steps. First, in order to avoid for a malicious node to create multiple sub instances and potentially gain control over the decision-making process (Sybil attack), a PoW is required for each node requesting to join the network in order to obtain an identity. This sets a minimum of computational resources required to join the collaborative network. Additionally, at the time of joining the network, the latency and bandwidth of the connection with the new member is estimated in order to have a global estimation of the peer-to-peer network capacity. Second, robots share information in a peer-to-peer manner outside of the blockchain network, but utilize the blockchain to record each of these transactions. With the initial PoW required to join the network, all robots can already have an estimation of the newly joined robot's available computational resources based on the time it needed to solve the PoW. This initial estimation might be very inaccurate, as multiple factors might significantly affect the total time required to solve a PoW puzzle, from the nature of the puzzle itself to the execution of other processes within the robot, and including delays introduced by the peer-to-peer network. Nonetheless, by utilizing periodic PoW solving times, and taking into account partial proofs of work also, a more accurate estimation can be obtained. In addition, intrinsic network overhead and other factors that can affect the solving time are mostly common to all robots in the network when averaging across a large number of solved puzzles. In consequence, these factors do not affect the decision making as they affect all robots in a similar way. Third, when data is exchanged between robots, map matching or data correlation methods are utilized in order to estimate the quality of the data. This is done through the submission in the blockchain of a data sample, or data stamp, whenever a data exchange between robots occur. This stamp is stored in the blockchain and can be compared by other robots that operate in the same environment in the future. We utilize the term data quality to refer to the accuracy and density of the information. In a mapping application, for instance, a 3D lidar is often able to produce more accurate and higher density data than an automotive radar would do in the same environment. Fourth, the quality of data calculated by the different robots is utilize to rank these across the network with a distributed consensus mechanism. Finally, the computational capabilities of robots receiving data and the ranking of robots generating data, where a robot can have one or both roles, are combined in order to distribute the peer-to-peer network bandwidth and optimize its usage.

The estimation of available computational resources at other robots is integrated into the PoW. However, the evaluation and ranking of the data provided by each of the robots requires a different data processing flow that needs to be integrated within the blockchain. When blockchains are utilized for a wider array of applications other than having a distributed ledger, smart contracts, which were defined by \cite{wood2014ethereum} as part of the Ethereum yellow paper and implemented within the Ethereum blockchain, are a key component in order to enable distributed execution of multiple applications. If a blockchain is utilized in order to manage the collaboration within a heterogeneous swarm of autonomous robots, scalability and low-latency transaction validation are essential. While sharding, when combined with other techniques as in the case of OmniLedger, enables linear scalability, the integration of smart contracts with sharding is not straightforward. \cite{kokoris2018omniledger} introduced Atomix, a client-driven lock/unlock protocol, to ensure that a single transaction can be committed across multiple shards, while enabling the possibility of unlocking rejected transaction proofs in specific shards. The original Atomix state machine can be extended to accommodate the execution of smart contracts across shards. In this paper, we propose the utilization of smart contracts in order to perform a collective and distributed evaluation and ranking of the data provided by the different robots. The operations required are paralellizable and can thus be executed across shards without introducing delays because no intermediate states need to be synchronized within a single evaluation. Therefore, we can ensure the scalability of the proposed approach and its applicability to real-time collaboration in swarm robotic systems.

The main contributions of this paper are: (i) the definition of a method for online and automatic estimation of available computational resources of all collaborating robots with a Proof of Work system; (ii) the definition of a strategy for ranking the quality of data provided by the different robots, (iii) an strategy for sharing the network bandwidth based on the previous two factors, i.e., the quality of data and the capabilities of different robots to process it in real time, and (iv) initial results that show the viability and present implementation possibilities of the proposed concepts. With these definitions, we aim at enabling effective collaboration in ad hoc robotic swarms with zero knowledge. In other words, robots are able to improve the way they share information with each other in a collaborative effort, without sharing explicit information about their identity, sensor suite or computational resources.

The remainder of this paper is organized as follows. Section 2 introduces key concepts of Blockchain technology that are utilized in our proposed system architecture. Then, Section 3 introduces strategies to manage collaboration through Blockchain technology in heterogeneous robotic swarms, in terms of data quality, trustability, management of bandwidth across the peer-to-peer network, and automated decision making of the data type and amount that is transferred to peers. Section 4 reports on initial results that show the viability of the proposed ideas. In Section 5, we discuss on the challenges and opportunities of the presented methods. Section 5 concludes the work and outlines future work directions.

\section{Background}

In this section, we introduce background concepts and theory. We focus on the concepts of consensus and, in particular, Proof of Work (PoW) and Proof of Stake (PoS), smart contracts, and scalability.

\subsection{Consensus}

Consensus mechanisms in a distributed system or decentralized network are those algorithms that allow agents to reach an agreement with respect to certain values, transactions or parameters whenever it is needed. Consensus mechanisms allow nodes in the network to trust others. The four most popular consensus mechanisms, according to \cite{li2017survey}, are Proof of Work (PoW), Proof of Stake (PoS), Practical Byzantine Fault Tolerance (PBFT) and Delegated Proof of Stake (DPoS), with other significant approaches including Proof of Authority (PoA), Proof of Elapsed Time (PoET) or Proof of Bandwidth (PoB). Both Bitcoin and Ethereum, the two most popular blockchains, utilize PoW, while Ethereum is shifting to a PoS-based consensus mechanism in Ethereum 2.0.

\subsubsection{Proof of Work}

The consensus mechanism introduced by Nakamoto~\cite{nakamoto2008bitcoin} as part of the Bitcoin was the Proof of Work (PoW), in what is the first known implementation of this mechanism as a way of achieving consensus in a distributed system. Firstly introduced by Dwork et al.~\cite{dwork1992pricing} as a filter to minimize the spamming capability of malicious email senders, the original idea of a PoW system has been invariant: to require a node in a network to solve a moderately computational intensive cryptographic problem in order to be able to make a transaction in the network, or validate its identity. In other words, a PoW is a cryptographic puzzle that is difficult to solve, while still being possible within a certain time and given certain hardware resources, and that it is very easy to validate. Thus, it takes a node a large amount of computing power to solve a PoW puzzle, but it takes other nodes in the network a small amount of computational resources to validate the solution.

In the Bitcoin and many successive blockchain systems, including Ethereum~\cite{wood2014ethereum}, a PoW-based consensus is introduced in order to validate new blocks in the blockchain. A block can be roughly described as a single entry in the distributed ledger containing information about a set of transactions. Each of the transactions in the block is confirmed when the block is validated by the network. In order to validate or \textit{mine} the data in a block, or body, a header is generated by hashing the information in the block body. The PoW consists on adding an extra cell to the block body, a random number denominated nonce, such that the block header meets some predefined conditions. These conditions are set on a block header's target hash as a function of the hashes of previous blocks. Once a node finds a solution to the PoW puzzle, i.e., a nonce with which the target hash is obtained, then it broadcasts the block to the rest of the network. Other nodes can easily validate the proposed solution, and then start mining a new block. The miner that solves the PoW for a given block obtains a reward in the form of newly created, or mined, cryptocurrency. When a block is mined by solving the corresponding PoW puzzle, it is added to the blockchain and the transactions that it contains are considered validated.

A security concern arises when two nodes find a nonce at the same time or when a second node, which has not received the proof yet, starts broadcasting the solution as well. In Bitcoin, nodes accept the previous block for which they receive a proof. In the case of two near-simultaneous proofs, the blockchain separates in two branches, or forks. Nakamoto~\cite{nakamoto2008bitcoin} introduced a rule in which the fork becoming longer or accumulating more mining difficulty would be judged as the authentic one by the network. 

The estimation of computing resources relying on Proof of Work has been studied earlier by Eyal et al.~\cite{eyal2015miner} and others~\cite{zhu2018survey, schrijvers2016incentive}. In his work, Eyal describes how not only full PoW solutions but also partial solutions can be utilized to ensure members in a mining pool do contribute to the collective mining effort. A mining pool is an association of nodes in a blockchain that utilizes PoW consensus in order to increase the probability of solving the PoW problem first, and therefore obtaining the corresponding reward. Nonetheless, if rewards are shared equally across the pool, then a malicious node might join the pool but never share its PoW solutions. The solution proposed by Eyal et al.~\cite{eyal2015miner} is to ask all nodes in the pool to share partial proofs, which can be validated by the rest of the pool members, and utilized to quantify the effort or computational power that nodes are dedicating, in average, to the mining task. A key aspect to take into account, however, is the distribution of complexity of the partial proofs of work as described in~\cite{eyal2015miner}. This helps avoiding that a malicious node always submits a partial proof of work with some minimum complexity but never tries to calculate the full proof. In a similar direction but with an opposite approach, Miller et al.~\cite{miller2015nonoutsourceable} proposed a mechanism for the definition of non-outsourceable puzzles which would discourage miners from joining mining pools or creating mining coalitions due to the inability of the pool members to estimate the partial progress of each individual node.

\subsubsection{Proof of Stake}

While PoW introduces a secure consensus mechanism in distributed networks, the increasing amount of computational resources not only limits the options of new nodes in the network to obtain mined currency, but also limits the maximum amount of transactions that can be processed as it takes an average time of 10 minutes to validate a block and all the transactions it includes~\cite{barber2012bitter}. In order to reduce the transaction validation latency, a Proof of Stake (PoS) mechanism was first implemented within the Nxtcoin~\cite{popov2016probabilistic, nxt2018whitepaper}. A PoS mechanism chooses the block validator based on the stake of different nodes, assigning a probability of being validator that is directly proportional to the amount of coins that a miner owns. A similar approach was introduced by Bentov et al.~\cite{bentov2016cryptocurrencies}, where the pure stake a miner owns is directly related to the probability of the miner to mine a new block, while the exact chance is calculated also based on the state of the current block.

A direct consequence of a PoS-based consensus is that nodes accumulating large amounts of cryptocurrency have a higher change of mining new currency, additionally increasing their stake. Moreover, the network becomes more vulnerable to attacks from these nodes or coalitions of nodes with large stakes. In Ethereum 2.0, a PoS consensus mechanism will be introduced within the so-called Casper protocol~\cite{buterin2019incentives}. Nonetheless, rather than considering the stake a miner owns, miners are required to put part of their coins at stake and lock them in a virtual safe during the validation process. Keeping the coins in a wallet without putting them at stake is not considered sufficient to be elected a validator in this case. In consequence, miners are incentivized to act in a honest manner as they risk losing all the coins they put at stake if faulty transactions are detected in their validated blocks. The mechanism through which a node loses the coins at stake because of faulty transactions are detected is denominated slashing.

Another clear benefit of PoS over a PoW mechanism is the much lower computational complexity of the operations involved, thus having a much smaller footprint in terms of energy consumption and computational resources required. 





\subsection{Smart Contracts}

The Ethereum blockchain~\cite{wood2014ethereum} introduced one of the most notorious smart contract platforms based on blockchain, by providing a Turing complete language as part of its framework~\cite{hildenbrandt2018kevm}. Ethereum introduced Solidity~\cite{solidity} as a language to implement smart contracts. In Solidity, a smart contract can be seen as a set of code instructions, or functions, and a set of initial, intermediate and final states (data). Both the data and the code resides at a specific address within the Ethereum blockchain. Smart contracts are part of the Ethereum Virtual Machine (EVM)~\cite{dannen2017introducing}. The EVM is a completely isolated environment for executing smart contracts within Ethereum, with no access to other processes, network connectivity or files in the system. In addition, the way smart contracts can access data from other smart contracts is also limited. 

\subsection{Sharding}

One of the main disadvantages of the consensus algorithms presented above, specially PoW due the large computational resources required, is scalability~\cite{vujivcic2018blockchain}. While Bitcoin only requires one broadcast per block, PBFT is based on multicast messages and also suffers from scalability in terms of communication cost~\cite{vukolic2015quest}. In all three cases, nonetheless, security does increase as the network becomes larger. Luu et al.~\cite{luu2016secure} presented Elastico in order to overcome the scalability limitations of previous blockchains, with an approach in which the network is partitioned into a set of smaller subnetworks or committees, also called shards. The definition of such committees is referred to as sharding, and Elastico was the first implementation of a sharding protocol for permisionless blockchains that is able to tolerate a predefined fraction of byzantine nodes in the network. 


\subsection{Scalability}

The main goals of Ethereum 2.0 can be summarized in five items: decentralization (to allow single-shard or system-level validation with consumer off-the-shelf hardware), resilience (to maintain operational conditions through network partitions and even if a significant fraction of the network goes offline), security (to deploy advanced crytographic strategies that enable a large-scale participation and validation), simplicity (to keep the consensus layer and top-level definitions as simple as possible), and longevity (to utilize either quantum secure components and mechanisms, or design the system in a way that it can be easily updated when possible for quantum secure equivalents), according to the Ethereum 2.0 specification~\cite{ethereum20specification, ethereum2017goals}. The roadmap to Ethereum 2.0 includes only a basic sharding approach in its initial phase, with no support for the EVM~\cite{ethereum20serenity}. The key innovations in the first phase of the Ethereum 2.0 deployment will be the utilization of the beacon chain with a PoS-based consensus mechanism~\cite{state1, state2}.

Some of the main ideas of Ethereum's solution in order to enable cross-shard communication, and the distribution of validators or shards for single-shard nodes, are the following~\cite{ethereum2018sharding}. First, the introduction of receipts, which are objects that are not stored in the shard's state but are defined in a way that Merkle proofs of their existence can be generated. This is useful for the simplest case where a larger number of applications have each a reduced number of users and do not need to share data often, so that each application can be contained within a shard and utilize receipts to communicate with other applications. Second, to offer transparent sharding, i.e., to dynamically create, merge or divide shards without the need for an application to be aware. Therefore, the sharding process is transparent to developers and they do not need to take sharding into account when defining smart contracts for different applications. Third, a solution for asynchronous cross-shard communication where receipts could be generated in order to revert transactions if necessary. If the system is biased such that reverts propagate faster than cross-shard requests, then this can effectively solve the problem of asynchronous cross-shard communication. Fourth, a strategy to avoid that an attacker sends multiple cross-shard requests from within a single shard. A proposed approach is to require an application that makes a cross-shard call to pre-purchase an amount of gas at the receiving shard (where the pre-purchase transaction occurs), which would be set as congealed. The amount of gas that can be congealed in a single shard is predefined, thus setting a limit to the amount of calls that can be made from other shards. Congealing gas avoids issues with volatile gas prices. In addition, a demurrage rate is included, such that the congealed gas is lost at a preset rate if it is not used within a receipt. Finally, congealed gas has the potential to be used for reliable intra-shard scheduling, even if only for the short term.

\section{Consensus in Collaborative Swarms with Blockchain}

In this section, we define an strategy for managing collaboration and establishing consensus in a collaborative robotic swarm utilizing a blockchain. Furthermore, we discuss how different aspects of the blockchain could be adapted to the specifics of robotic cooperation, where the most valuable token that can be exchanged between robots is data. Therefore, part of the security focus is shifted from the transaction validation point of view to the data quality aspect. In that regard, a blockchain can be utilized to establish a secure way of evaluating and ranking the quality of data provided by the different robots.

The approach presented in this paper can be generalized towards achieving consensus in a large multi-robot system, or swarm. Nonetheless, we focus on a specific problem that is particularly significant: cooperative mapping and collaborative sensing or perception within a heterogeneous team of autonomous robots operating in the same environment. With perception, localization and mapping being three of the cornerstones behind fully autonomous operation, the problem of collaborative sensing for enhancing the situational awareness of each of the individual robots sets the basics towards more complex collaboration.

We focus on heterogeneous robotic systems because of their dynamism and the wider variety of applications that they enable. Heterogeneous multi-robot systems or larger swarms have been studied for over two decades~\cite{arkin1997cooperative, quinn2003evolving, tuci2018cooperative, zhao2015evolved, akbari2019novel}. In addition, we consider ad hoc swarms where the number of robots can change over time, and their properties are not within a predefined set. These changes present multiple challenges from the perspective of a heterogeneous resources management system. In a homogeneous ad hoc robotic system, different parameters such as computation power, bandwidth distribution or type and amount of data to be shared can be either predefined or calculated based on a preconfigured strategy. This means that, in most cases, the way that robots interact with each other will not suffer from sudden changes. However, the same does not apply to heterogeneous robotic systems, where a new robot joining the collaboration effort might have a sensor suite or computational resources very different from the rest of collaborating robots. In that case, all the robots need to adapt their collaboration schemes, with potentially significant changes in the way information flows within the network and in the selection of robots that have priority over others to either share or receive information.

An additional assumption that we make is that robots can be anonymous and that all robots have the same role within the blockchain. However, some conditions must be set in this regard. As we are utilizing a blockchain in order to manage the collaboration and consensus between the different individual robots, a minimum number of nodes must remain in the network in order to keep the blockchain alive. Alternatively, infrastructure in the operation environment can be utilized in order to provide the backbone of the blockchain, making sure that it stays alive and with a minimum level of security so that previously stored data can still be trusted or utilized by new robots joining the collaborative network.

Keeping the blockchain alive and all previous records has the disadvantage of a higher overhead when a robot joins the network. Nonetheless, in this paper we present an approach to evaluate the quality of data provided by the different robots and utilize a ranking based on this metric in order to manage the collaboration. In the case of having robots operating only at sparse time intervals, with long idle periods in which the robots are off or offline, these might lose their status and need to regain it every time they join the network. This has a negative impact not only in the robot itself, which would need to regain the trust of its peers in terms of data quality, but also for the rest of robots, which might be receiving less accurate data until the previous status of the new collaborating peer is regained. Therefore, the level of optimality of the collaboration might be reduced. In general, there is a trade-off between the benefits that the data history stored in the blockchain brings to the collaboration with the drawbacks in terms of synchronization and overhead when robots join or leave the network. It is left as a design decision up whether only the blocks generated in a certain recent time frame are downloaded by new robots or the whole blockchain is. Not having the complete blockchain is not an issue in terms of transaction validation because we introduce a demurrage effect inspired by the design ideas of Ethereum 2.0 to the network's cryptocurrency. Therefore, the tokens that individual robots had in their stake earlier than a certain time threshold are no longer usable.

Having a connected infrastructure that supports the blockchain is an interesting approach that has the potential to enable multiple applications in the era of 5G and beyond connectivity. With network slicing, softwarization and virtualization, and the ability to support a variety of different verticals, blockchains and other systems could be deployed within the base stations. In those cases, permissioned blockchains might be considered a more suitable solution, where the role of the infrastructure could be more related to validating and increasing the level of security. In any case, in this paper we also consider fully distributed and anonymous permissionless blockchains. A more in depth analysis on this topic is given within the discussion of the design principles.

We focus on providing a strategy with potential to solve two existing challenges in heterogeneous multi-robot collaboration: (1) the management of bandwidth in a peer-to-peer network between different robots, potentially having a myriad of sensing capabilities and computational resources, choosing which robots have priority to transmit their data and which robots will receive specific data batches; and (2) the decision making in terms of what specific data do robots share in order to optimize the benefits that results from the collaboration process. Both concepts are closely related; if a specific robot is able to obtain higher quality data, then a more optimal collaboration can be achieved if its data is shared among the interested peers. In that case, this particular robot should have priority in terms of bandwidth usage. While this argument utilizes only the sensing capabilities of robots, a similar approach can be taken in terms of the computational resources of each robot. In a smart city environment, an autonomous delivery drone with limited sensing abilities, for example running visual inertial odometry with a single onboard camera, could benefit from the data extracted by a self-driving car equipped with multiple cameras, radars and a high-quality multi-channel 3D lidar. However, if the car simply streams its data, the drone would be barely able to process most of it due to constrained computational resources in its onboard computer. Therefore, it might either discard most of the data when receiving it, or accumulate it and induce a delay in the processing with the consequent latency increase in localization or mapping tasks.

In summary, the main difference of our approach with respect to previous works is that robots do not need to share information about their sensing and computational capabilities, yet they can be able to optimize the way they are collaborating. We understand collaboration optimality as the enhancement of each of the robot's situational awareness resulting from the analysis of data provided by their collaborating peers. By utilizing a blockchain framework, and adapting the latest developments in the blockchain field, we are able to provide robots with means for more efficient ways of collaboration without having to share, trust and interpret specific data about their sensors or data processing capabilities.

\subsection{Design Principles of a Blockchain for Multi-Robot Collaboration}

The utilization of a blockchain for managing the collaboration between robots, based on the assumptions and situations described above, can have multiple benefits. However, multiple design aspects that can significantly affect the way the blockchain is utilized have not been specified so far. It is not the aim of this paper to provide a specific set of methods for heterogeneous robotic collaboration, but instead to provide a set of basic strategies and design concepts that can be utilized towards the definition of advanced collaboration schemes in the future. In this section, we overview the different approaches that can be taken in different blockchain aspects. 

\subsubsection{Blockchain Genesis}

The first design decision is whether a longevous blockchain is preferred, or ad hoc blockchains are created when needed. Both options present significant challenges from the implementation point of view, and both can be utilized in different environments or specific application scenarios. We overview the main benefits and drawbacks of each of them. In future work, we will put into practice these two options and evaluate different parameters in order to provide more insight.

\subsubsection*{I. A Single Longevous Blockchain}
In the case of a single longevous blockchain with ad hoc collaboration where new robots might join the network, and other robots might go offline for long periods of time, the main challenge is to ensure that a minimum number of nodes is able to securely maintain the blockchain state. We believe that this type of blockchain can only be effectively implemented when connected infrastructure is added into the blockchain. Whether it is a private or semi-private blockchain in an industrial environment, or an open permisionless blockchain that robots can join to improve their operational performance in a smart city, there must be a minimum set of nodes that are fixed in the environment and are able to support the blockchain even in the case that no robots are active. Furthermore, connected infrastructure, such as mobile network access points or smart gateways near public Wi-Fi hotspots around a city, can be utilized to define a standard for communication within the blockchain and to publish its existence so that anyone can decide to join it.

An immediate question that raises when considering connected infrastructure, which naturally has a different role from robots joining the network, is whether a permisionless blockchain is the best option, or a permissioned consortium blockchain where the infrastructure nodes have a predefined role of validators is more secure. We favour the permissionless and open blockchain option because its benefits in terms of flexibility and because the key aspect of a blockchain for robotic collaboration does not rely as much on the transaction security as it rather does in the integrity and quality of shared data, which can not be validated with a traditional blockchain approach.

\subsubsection*{II. Ad Hoc Blockchains}
An alternative to a single blockchain, potentially supported by connected infrastructure, is to create and destroy collaborative networks on the fly. The main disadvantage of this approach is on how to define the conditions under which the blockchain is started, and which entities are allowed to initiate the process. This option is not suitable for applications where individual and anonymous robots are collaborating, with potentially independent developers. Instead, this might be a more suitable approach for situations where a single controller or developer is deploying a large robotic swarm, which may already include a collaboration scheme or not. In that case, the blockchain genesis can be established by the developer, and its existence made public, opening the door to other robots or swarms to join and share data. In this case, the newly joined members should put trust on the blockchain initiators.

A similar situation where an ad hoc blockchain could be applied is in the case of multiple end-devices, vehicles or robots being produced by a single manufacturer but utilized by different individuals. For instance, this could be the case of a company selling self-driving cars. These could be preconfigured to automatically detect other cars from the same manufacturer in the vicinity. In the event of multiple cars converging in a near area, a blockchain could be started without the car owners being aware of it, and these could start benefiting from the data collected at other vehicles, with enhanced autonomous operation. It would be then a decision of the manufacturer whether to broadcast the existence of this network for other vehicles to connect or not. In comparison with a direct cooperative data sharing approach, the utilization of the blockchain and the data quality validation strategies presented in this paper would ensure that faulty sensors or tampered sensor data can be detected by the network if certain conditions are met under which vehicles are able to validate each other's data.

\subsubsection{Consensus Mechanism}
While PoW shows an additional potential other than providing a means for consensus, its many drawbacks described in Section 2 make it unsuitable for long-term scalability. In consequence, we propose the utilization of a Proof-of-Stake system for validating transactions, while maintaining a periodic PoW for computational resource estimation. In any case, the energy and time spent by miners towards the PoW is not futile, as it will be utilized by the network to estimate the available computational resources at different nodes. In general, we propose the utilization of a protocol similar to the Casper PoS protocol in Ethereum 2.0, while a PoW is required to be executed periodically for nodes to be still considered part of the peer-to-peer network \cite{buterin2017casper, ethereum20specification}.

In a traditional blockchain architecture, a transaction is validated when the corresponding block where it has been included is mined. In our case, transactions in the proposed network architecture are data exchanges between one robot and a subset of its peers. The data itself is shared within the peer-to-peer network but outside of the blockchain. A certain sample is shared within the blockchain as part of the transaction body so that the whole network can run the data quality evaluation and ranking procedure. Robots do not wait for transactions to be validated within a block before sharing the data. Rather than coining new cryptocurrency though the consensus mechanism, we propose the utilization of of the periodic PoW mechanism to provide robots with a preconfigured and fixed amount of cryptocurrency. While in the case of the Ethereum blockchain the Ether can be utilized both to perform trade between peers, and to purchase the network's execution time, in this paper we only consider the latter scenario. Therefore, the only use of the cryptocurrency is to be able to perform data transactions and be sure that data will be forwarded through the peer-to-peer network.

Rather than providing rewards for block mining as an incentive, the PoW puzzles are compulsory for nodes to be considered part of the network. The time between PoW requests can be preset within the network configuration, as well as the procedure by which the next PoW is automatically calculated based on the blockchain's state or other parameters. In order to ensure that nodes do not collect a large amount of coins, we propose the implementation of a demurrage mechanism as in Ethereum 2.0, where all the cryptocurrency is effectively congealed and disappears within a certain time interval. In addition, penalties must be included to further control the utilization of cryptocurrencies. In doing so, we can limit the amount of data that dishonest nodes are able to send over the peer-to-peer network.

\vspace{0.5cm}
\subsubsection{Security Concerns}

In a robotic collaboration system, data exchanges are the most valuable tokens. However, including all data in the blockchain would significantly reduce its usability because of the impact on scalability and the latency that such amount of data processing and validation in all nodes would induce. Therefore, in order to provide basic means for the collaborating robots to decide the level of trust that they put into a certain robot, data samples are submitted to the blockchain and evaluated within the network through smart contracts. These samples are then ranked and utilized in order to estimate whether a robot is honest or not, but also which robot is able to provide more accurate or useful data given a particular request.

It is not within the scope of the collaborative decision making presented in this paper the definition of how the data quality ranking is taken into account at each individual robots. This is for application developers and must be implemented separately at each robot controller.

\subsubsection{Scalability}

In order to ensure efficient scalability of the proposed blockchain architecture, we propose the utilization of spatial shards for local decision making in terms of data quality evaluation. Therefore, both a local ranking and a global ranking are kept in record at the shard chains and global chain. This is a useful approach in order to reduce the network load and induced latency. The consensus mechanisms would run in parallel shards which would be defined based on the Ethereum 2.0 standards. A single validator thus belongs to two kinds of shards, spatial shards utilized for running local smart contracts regarding local data quality ranking, and randomized shards running the consensus mechanism and maintaining a global ranking with separate smart contracts. More insight into the definition of the local and global rankings is given in the data evaluation section.

\subsection{PoW for Online Estimation of Computational Resources}

In this section, we propose a methodology for estimating the computational resources of each of the collaborating robots by exploiting the PoW puzzles utilized in the blockchain in order to validate blocks. The time required to solve a PoW puzzle can be utilized as an indicator of the available computational resources at a given robot, and partial proofs can be used in case robots are not able to solve a PoW puzzle within a certain predefined time interval. We utilize the term available, rather than total computational resources, because we assume that robots are able to operate autonomously on their own, and utilize the collaboration in order to improve the accuracy of the different methods that they already run. Therefore, robots must decide which amount of resources do they want to reserve for the collaborative effort; the more resources they put into solving PoW puzzles, the more data they are able to obtain, as the amount of data is calculated based on the available processing power in order not to overload the receiving robot with more data that it is able to process.

In a typical PoW utilization for block mining in a blockchain, once a miner finds a solution to the PoW puzzle and broadcasts it, all other miners automatically discard their solutions and start working on mining a new block. However, this can only give an idea about the processing power of the node that was able to mine the block. In order to be able to obtain useful information regarding all nodes in the network, partial proofs of work can provide more insight into the effort that different nodes put towards the PoW puzzles.

The utilization of partial proof of works has been previously been proposed in different mechanisms that secure and raise the level of fairness in mining pools \cite{eyal2015miner, eyal2018majority, zhu2018survey, rosenfeld2011analysis, schrijvers2016incentive}. Mining pools utilize various payout systems in order to distribute the mined coins between their miners even if individual nodes have not been able to provide a full PoW solution. For the estimation of computation in a robotic system, a simplistic approach is enough. One naive solution is, for instance, that each robot is assigned a different PoW puzzle with equivalent complexity. This helps to avoid two robots submitting the same partial or full PoW solution while only one is actually calculating it. In this case, we do not need to consider the presence of malicious nodes that put less computational resources towards solving PoW that they can. This is because the conclusion from the network would be that the processing power is more limited at those nodes, and other robots would therefore send less data. In this approach, solving PoW puzzles is not the means towards a monetary reward, but instead towards a data reward. The faster a PoW problem is solved, the more data a robot is likely to receive from its peers. Therefore, individual nodes would gain nothing and only incur in their own detriment by lying to the network with less complex partial PoW solutions. Because the result of the PoW has to be shared with the rest of robots and can be easily validated by each of them, robots cannot provide fraudulent data regarding their computing capabilities. The reason robots might perform such malicious actions could be because either they want to destabilize the network or they want to increment the amount of data they are receiving in a network with limited bandwidth where most robots might have extra processing power that is not being put into use.

The mining difficulty should be set to a fixed value, in contrast with Ethereum's adaptable PoW complexity, so that robots with lower computational resources can also be part of the collaborative sensing scheme. Nonetheless, the complexity should be enough to ensure that only robots with a minimum level of computational capabilities are able to participate. In a similar way, the network connectivity of the robots must be put to test before joining the network in order to avoid bottlenecks and dub-optimal collaboration. At the same time, if the PoW is too easy to solve, then the communication overhead might play a more significant role. In general, if the network-wide communication latency is at least one order of magnitude smaller than the minimum time required to solve a PoW puzzle, then a simple averaging could suffice for more accurate, long-term estimation. Nonetheless, the estimation should be able to adapt to changes in the available computational resources. This can happen because robots might be running other computation intensive processing algorithms that are only executed at certain intervals, or only when a series of events occur. In order to do this, a naive approach would be, for instance, to select the last $N$ PoW proofs or partial proofs such that for all (or most) $M<N$, the estimated computational capabilities $C_M$ and $C_N$ have low variance, i.e., $\mu_{C_M}-\sigma_{C_M} \leq \mu_{C_N} \leq \mu_{C_M} + \sigma_{C_M}$. However, potential outliers should be taken into account and a minimum number of partial PoW solutions $N>\varepsilon$ utilized in the collective estimation procedure. If the nature and capabilities of robots collaborating through the network changes, with significant increase or reduction of the average computing power, an alternative approach to adapting the PoW complexity is to set a maximum time that robots dedicate to the mining effort, even if none of them is able to produce a full PoW solution. In this case, however, the maximum latency in the peer-to-peer network should be taken into account in order to calculate the timeout interval, and its value should be negligible in comparison.

\subsection{Data Evaluation - Proof of Quality}

We are basing the collaboration management in two parameters: the estimation of the available computational resources and the evaluation and ranking of the quality of data provided by different robots. For the first aspect, we have proposed the utilization of a PoW scheme in order to maintain an online estimation during operation. For the second aspect, the main idea presented in this section is to share within the blockchain a data stamp, or sample, whenever a data exchange transaction between two robots, has been made. Thus, not all the data is stored in the blockchain, but can be transmitted through a direct connection, an external network or the peer-to-peer network. We assume that the only connection between robots is the peer-to-peer network. This matches with the assumption that robots are anonymous, and therefore they do not necessarily have any means of contacting their peers. The main argument behind this assumption is that collective decision making in terms of what data is shared between certain pairs of robots is strongly affected by the constraints of bandwidth or latency inherent to the peer-to-peer network. In external channels exist, it is not straightforward to consider them within this distributed process.

The cryptocurrency in the proposed collaborative network has no real value. Instead, the most valuable asset is data. Therefore, we put the focus around the data and how to evaluate and rank its quality. The following parameters are utilized to evaluate the data stamp: (i) the type of data that has been provided; (ii) the density or resolution of the data sample; and (iii) the comparison of this data sample with historical samples from other robots that are or have been operating in a close location. Regarding the type of data that is being shared, it can be classified, for instance, into visual data (camera images), point cloud data (from lidars, depth cameras), or radar data. Then, the resolution of these images, or the density of a lidar point cloud is also taken into account. Based on this evaluation, penalties can be defined for nodes that fail to provide certifiable data stamps by providing a reduced share of newly coined cryptocurrency.

A ranking of the quality of data is not kept within the blockchain. This is because the type of data can only be evaluated based on the global needs of the system, which can considerable change over time. What is stored in the blockchain is the results of comparison of data stamps from the same environment. This result can be (i) a confirmation that the data matches, with no ability to provide further information; (ii) a confirmation that the data matches and that the current sample is either less dense and included in the sample, or that the new sample has more resolution; or (iii) a mismatch event where the robot has been unable to confirm that both the historical sample and its new sample represent the same object or environment.

A trustability concern arises when considering the event of a subset of nodes submitting bogus data stamps to the blockchain, which are in turn validated by other robots in an attack coalition. However, in order to do this, the attacker coalition must hold a majority in a given spatial localization. We propose the utilization of a validation graph, where two nodes are connected if any one has validated the other's data stamp, and its analysis in order to detect fully connected or almost fully connected subgraphs, or whether it is a disconnected graph with multiple separated components. While this can give an idea of trust, the decision-making process from this information is not straightforward. Consider for instance the case in which a certain group of honest robots are sharing and validating each other's real data, and an equivalent number malicious nodes is sharing and validating counterfeit data. If these are the only collaborating robots in the vicinity, then for a robot in another location it is impossible to discern between them. However, this might not necessarily be a problem if no other robot is utilizing data from that particular location. In the long-term, given a majority of honest collaborating robots, and assuming that most locations are visited by a large enough number of the honest robots, then the counterfeit data stamps will be eventually detected and the set of malicious nodes will be labelled as dishonest.

\subsection{Peer-to-Peer Data Sharing Scheme}

So far we have provided an approach that assumes that robots do not share explicit information about their sensors, or the on-board hardware resources. This means that the maximum level of optimality that can result from such collaboration is limited by how well can the different proposed approaches abstract and model the robots' resources and capabilities. In other words, if the estimation of computational resources that is obtained via PoW cannot be performed in an accurate manner, this inherently limits the maximum level of performance of the proposed system. A similar situation occurs with the estimation of data quality and robot trustability level.

If robots decide to share data without filling a transaction within the blockchain, it must occur either through a direct link or outside of the peer-to-peer network. If a third node receives a request to forward data between a given pair of robots, it only does so if the corresponding hash of the data sample has been already included in the blockchain.

In the same sense than in Ethereum nodes buy gas in order to execute smart contracts, each transaction consumes a given amount of cryptocurrency. In order to avoid a situation in which a malicious node would start multiple transactions, effectively double-spending its stake, sending large amounts of data in order to collapse the network, we propose that a strategy similar to the gas congealing scheme in Ethereum 2.0 can be utilized. The key difference is that all the cryptocurrency that a robot has in its stake is congealed and subject to the demurrage effect.

The decision making in terms of the data to be transmitted can be established as an optimization problem where the bandwidth of the peer-to-peer network and the available computational resources at the nodes receiving the data are considered constraints. The function to be maximized is a weighted sum of the data that robots receive based on their requests, and a measure of the trust put into the quality of that data. In this paper, we provide a high-level approach and do not delve into the details of specific calculations, which will be considered in future work.

Consider that at a certain time instant the blockchain is formed by a coalition of $N$ nodes, with their indexes represented by the set $[N]={0,\dots,N}$. A new PoW problem is considered by all robots, where $L_H$ represents the size of the PoW hash in bits, and $PoW_i \:\forall i < N$ is the full or partial solution that robots provide and is verifiable by their peers, $PoW_i\in\mathbb{Z}_2^{L_H}$. The available processing power at each robot is then estimated relatively to that of other robots, and mapped to the interval $\mathbb{R}^{[0,1]} = \{x\in\mathbb{R} \vert x \in [0,1]\}$. We denote the estimator with $\hat{\mathcal{C}}:\mathbb{Z}_2^{L_H N} \Longrightarrow \mathbb{R}^{N[0,1]}$ which takes as input the set $\{PoW_i\}_{i\in[N]}$ and outputs the set $\{\hat{\mathcal{C}}_i\}_{i\in[N]}$. During the estimation calculation, an additional parameter $\mathbb{D}_{max}$ is calculated, which defines the maximum amount of data that the robot with the most computing power is able to process per second, in bytes. Upon submitting a PoW full or partial solution, each robot $i$ also submits a data request, denoted $DR_i$, which contains a requested amount, in bytes, type of data (image, point cloud, radar or others) and a minimum and maximum resolution or density $DR_{i,j} = struct\{type, \: max\_size, \: min\_res, \: max\_res\}$, where $j$ varies for each different data type. Additionally, robots submit information about the data they are able to provide, or available data, with the same type of information $AD_{i,j} = struct\{type, \: max\_size, \: min\_res, \: max\_res\}$, where again $j$ iterates over the available data types. We suppose that an error function $\mathcal{E}_{DR_{i,j},\:DA_{k,j}}>0$ is given that increases as the mismatch between the desired data size and resolution in a data request from robot $i$ and the available data at robot $k$. Finally, robots share their position in a global reference frame and an estimation of its error. This is utilized in order to divide the robots in spatially-defined shards, and at individual robots to decide how they utilize the received data.

We assume that the data quality evaluation provides a value $\mathcal{Q}_i\neq0\in\mathbb{R}$ that can be negative and represents the trust that the network puts on robot $i$. We do not provide a specific formula to calculate this value, but instead refer to the guidelines described in the previous section. Given the maximum bandwidth of the link between robots $i$ and $j$ in the peer to peer network, $BW_{i,j}$, we can now formulate the optimization problem that is solved as part of a smart contract deployed in the network in order to make decisions with respect to the data that is shared between robots:

\begin{equation*}
\begin{array}{rl}
    \text{arg}\displaystyle\max_{X}  &\quad  f(X) = \displaystyle\sum_{i=1}^N \left( \displaystyle\sum_{j\vert x_{ij}\in X} \alpha\mathcal{Q}_j + \beta \dfrac{1}{\mathcal{E}(x_{ij})} \right) \\[+23pt]
    \begin{array}{l} \text{subject to: }\\ \: \end{array} & \begin{array}{rl}\quad  x_{ij,\:size}   \leq \mathbb{D}_{max} \cdot \hat{\mathbb{C}}_i   &\quad \forall x_{ij} \in X\\
                  x_{ij,\:size}   \leq  BW_{i,j}                                   &\quad \forall x_{ij} \in X \end{array}\\
\end{array}
\end{equation*}

where $X=\{x_{ij}\}$ represents a data exchange between the pair $(i,j)$ with a given size $x_{ij,\:size}$, and $\mathcal{E}(x_{ij})=\mathcal{E}_{DR_{i,j},\:DA_{k,j}}$. The parameters $\alpha$ and $\beta$ define a weighted sum between data usability and data trustability that must be set depending on the range than the data quality and data error match functions can give. We have considered that the bandwidth $BW_{i,j}$ of the link between robots $i,j$ is independent of data that might travel between the same link but does have a different recipient. This is a unrealistic assumption in a peer-to-peer network that potentially relies on a mesh network for communication. However, that is a problem within the graph theory domain and we do not consider within the scope of this paper.

\section{Results}

In this section, we partly evaluate the potential of the proposed methods. In particular, we focus on studying the correlation between the computational resources required to solve a PoW problem with the execution of different algorithms widely utilized in robotics. In addition, we show different data samples that could be utilized within the data evaluation scheme. 

\begin{table*}
    \centering
    \caption{Hashing power of different boards typically utilized as onboard computers in robotics (in hashes per second). The hashing algorithm was SHA256 and the tests involved solving PoW puzzles with 22, 23 and 24 bits of difficulty. The hashing puzzles were solved running within a single thread. The standard deviation shows \textit{NA} when it is below 1000~hashes/second.}
    \small
    \begin{tabular}{@{}lccccc@{}}
        \toprule
                                            & Intel Up  & Intel Up Gateway  & NVIDIA Jetson TX2    & Intel i5-6200 \\
        \midrule
        \textbf{Avg. Hashing power (h/s)}   & 89000     & 79000             & 184000        & 561000 \\
        \textbf{Std. Hashing power (h/s)}   & $NA$      & $NA$              & 1000          & 16000 \\
        \bottomrule
    \end{tabular}
    \label{tab:pow}
\end{table*}

\begin{table*}
    \centering
    \caption{Classification latency in tensorflow for a CNN classifying the CIFAR-10 dataset~\cite{krizhevsky2009learning}, and visual odometry (VO) latency for the Kitti dataset~\cite{geiger2013vision}. The standard deviation shows \textit{NA} when it is below 10$\mu s$.}
    \small
    \begin{tabular}{@{}lccccc@{}}
        \toprule
                                                    & Intel Up  & Intel Up Gateway  & NVIDIA Jetson TX2    & Intel i5-6200 \\
        \midrule
        \textbf{Avg. classification latency ($\mu s$)}      & 4400     & 5000              & 700         & 770 \\
        \textbf{Std. classification latency ($\mu s$)}      & 500      & $NA$              & 40          & 60 \\
        \midrule
        \textbf{Avg. VO latency ($ms$)}     & 200      & 210               & 108         & 59 \\
        \textbf{Std. VO latency ($ms$)}     & 112      & 119               & 50          & 29 \\
        \bottomrule
    \end{tabular}
    \label{tab:nn_vo}
\end{table*}

\subsection{PoW Metrics}

We have utilized four different computing platforms to evaluate the consistency of the relationship between the hashing power and different types of algorithms that autonomous robots might run during their operation. The PoW solver has been implemented a single-thread process so that it can then be run in parallel in order to take into account also the number of available threads or cores in the robot's onboard computer. The four computing platforms are (1) an Intel Up board, (2) an Intel Up Gateway, (3) an NVIDIA Jetson TX2, and (4) an Intel i5-6200U CPU. The NVIDIA Jetson TX2 has been specifically selected because it has an embedded Pascal GP. As the PoW relies only on a processor, it is not able to model accurately the processing power for applications that can be inherently accelerated with a GPU.

The hashing power of the different devices is shown in Table~\ref{tab:pow}. The number of hashes varies almost an order of magnitude between the least and most capable ones. The standard deviation of the hashing is below 3\% for all the devices under test. Table~\ref{tab:nn_vo} then shows the latency of two different types of data processing processes: a convolutional neural network (CNN) classification for the CIFAR-10 dataset~\cite{krizhevsky2009learning}, and visual odometry for the Kitti dataset~\cite{geiger2013vision}. The standard deviation in the case of the CNN classification latency is always lower than 12\%, while the visual odometry latency is more variable and has a standard deviation of over 50\% in some cases.

\begin{figure}
    \centering
    \setlength\figureheight{.35\textwidth}
    \setlength\figurewidth{0.49\textwidth}
    \small{
\begin{tikzpicture}

\definecolor{color0}{rgb}{0.886274509803922,0.290196078431373,0.2}

\begin{axis}[
axis background/.style={fill=white!89.80392156862746!black},
axis line style={white},
height=\figureheight,
tick align=outside,
tick pos=both,
width=\figurewidth,
x grid style={white},
xlabel={Computing platform},
xmajorgrids,
xmin=60, xmax=290,
xtick style={color=gray!66.66666666666666!black},
xtick={100,150,200,250},
xticklabel style = {rotate=60.0},
xticklabels={Up,UpGtw,TX2,i5-6200},
y grid style={white},
ylabel={Hashing power * odometry latency},
ymajorgrids,
ymin=4923.6, ymax=51484.4,
ytick style={color=gray!66.66666666666666!black}
]
\path [draw=color0, semithick]
(axis cs:100,7040)
--(axis cs:100,24960);

\path [draw=color0, semithick]
(axis cs:150,8099)
--(axis cs:150,29281);

\path [draw=color0, semithick]
(axis cs:200,10672)
--(axis cs:200,29072);

\path [draw=color0, semithick]
(axis cs:250,16830)
--(axis cs:250,49368);

\addplot [semithick, color0, mark=asterisk, mark size=3, mark options={solid}, only marks]
table {%
100 16000
150 18690
200 19872
250 33099
};
\end{axis}

\end{tikzpicture}}
    \caption{Relation between hashing power and visual odometry latency for the different computing platforms. The ratio is maintained mostly constant if the high variance is taken into account, due to the task being run completely on the processor.}
    \label{fig:vo_ratios}
\end{figure}
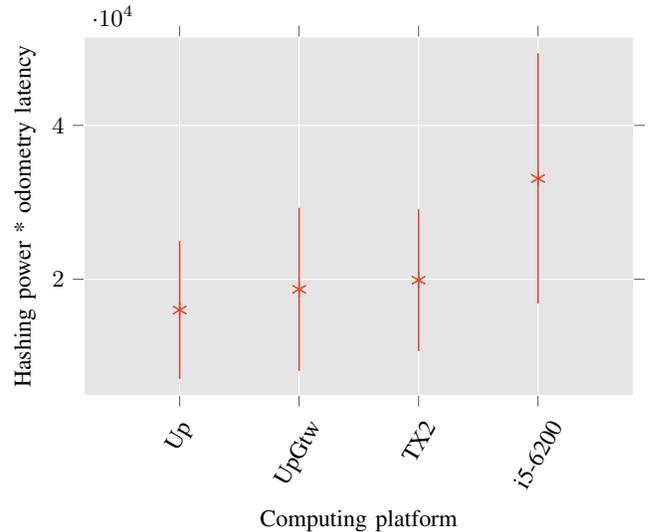

Figures~\ref{fig:nn_ratios} and \ref{fig:vo_ratios} show the relationship  between the hashing power and the performance of the classification and odometry algorithms, respectively, for the different devices. In the case of the CNN classification, the ratio is mostly constant, except for the NVIDIA Jetson TX2. This shows that the PoW puzzle can model the processor capabilities with high fidelity, but fails when other types of resources (GPU in this case) play a significant role. The CNN classification runs at lower latency on the NVIDIA Jetson TX2 than on the Intel i5 processor.

In a full system implementation, multiple PoW schemes should be designed to model the different types of computing resources that could be available within the collaborating robots. In future works, we will analyze different types of PoW puzzles and hashing algorithms, as well as strategies to avoid specialized accelerators such as ASICs.

\begin{figure}
    \centering
    \setlength\figureheight{.35\textwidth}
    \setlength\figurewidth{0.49\textwidth}
    \small{
\begin{tikzpicture}

\definecolor{color0}{rgb}{0.886274509803922,0.290196078431373,0.2}

\begin{axis}[
axis background/.style={fill=white!89.80392156862746!black},
axis line style={white},
height=\figureheight,
tick align=outside,
tick pos=both,
width=\figurewidth,
x grid style={white},
xlabel={Computing platform},
xmajorgrids,
xmin=60, xmax=290,
xtick style={color=gray!66.66666666666666!black},
xtick={100,150,200,250},
xticklabel style = {rotate=60.0},
xticklabels={Up,UpGtw,TX2,i5-6200},
y grid style={white},
ylabel={Hashing power * inference latency},
ymajorgrids,
ymin=104.2305, ymax=482.8395,
ytick style={color=gray!66.66666666666666!black}
]
\path [draw=color0, semithick]
(axis cs:100,399.2)
--(axis cs:100,400.8);

\path [draw=color0, semithick]
(axis cs:150,347.1)
--(axis cs:150,436.1);

\path [draw=color0, semithick]
(axis cs:200,121.44)
--(axis cs:200,136.16);

\path [draw=color0, semithick]
(axis cs:250,398.31)
--(axis cs:250,465.63);

\addplot [semithick, color0, mark=triangle*, mark size=3, mark options={solid}, only marks]
table {%
100 400
150 391.6
200 128.8
250 431.97
};
\end{axis}

\end{tikzpicture}}
    \caption{Relation between hashing power and classification latency for the different computing platforms. The ratio is mostly constant except for the NVIDIA Jetson TX2, which offers higher performance in the case of the CNN classification due to its integrated Pascal GPU.}
    \label{fig:nn_ratios}
\end{figure}
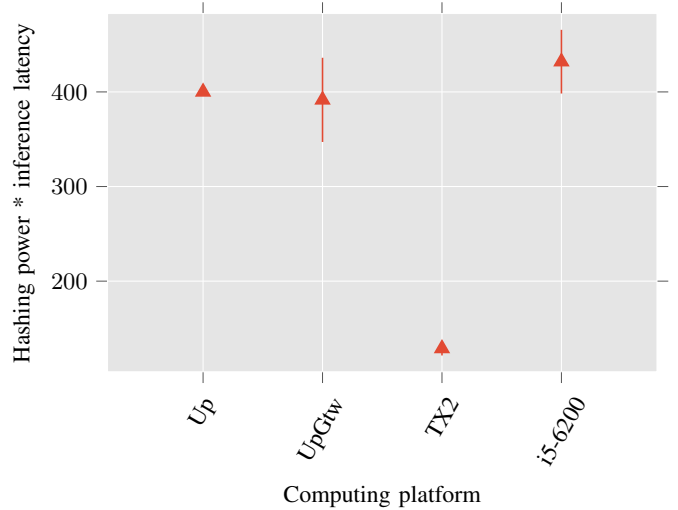

\subsection{Data Sharing Scheme}

For the data sharing scheme, we show the amount of data that would need to be shared within the blockchain to compare it between different vehicles. We also show how the data characterization can depend on the point of view or distance to the feature that is being utilized for the data sample. This is important to take into account when comparing the same feature from data samples submitted to the blockchain by different vehicles or robots at different times. Because the point of view of the object in the data sample and the distance to it might affect the properties of the data in the sample, a relation between these must be defined beforehand. Then, different data samples can be compared appropriately.

\begin{figure*}
    \centering
    \begin{minipage}{0.44\textwidth}
        \centering
        \includegraphics[width=\textwidth]{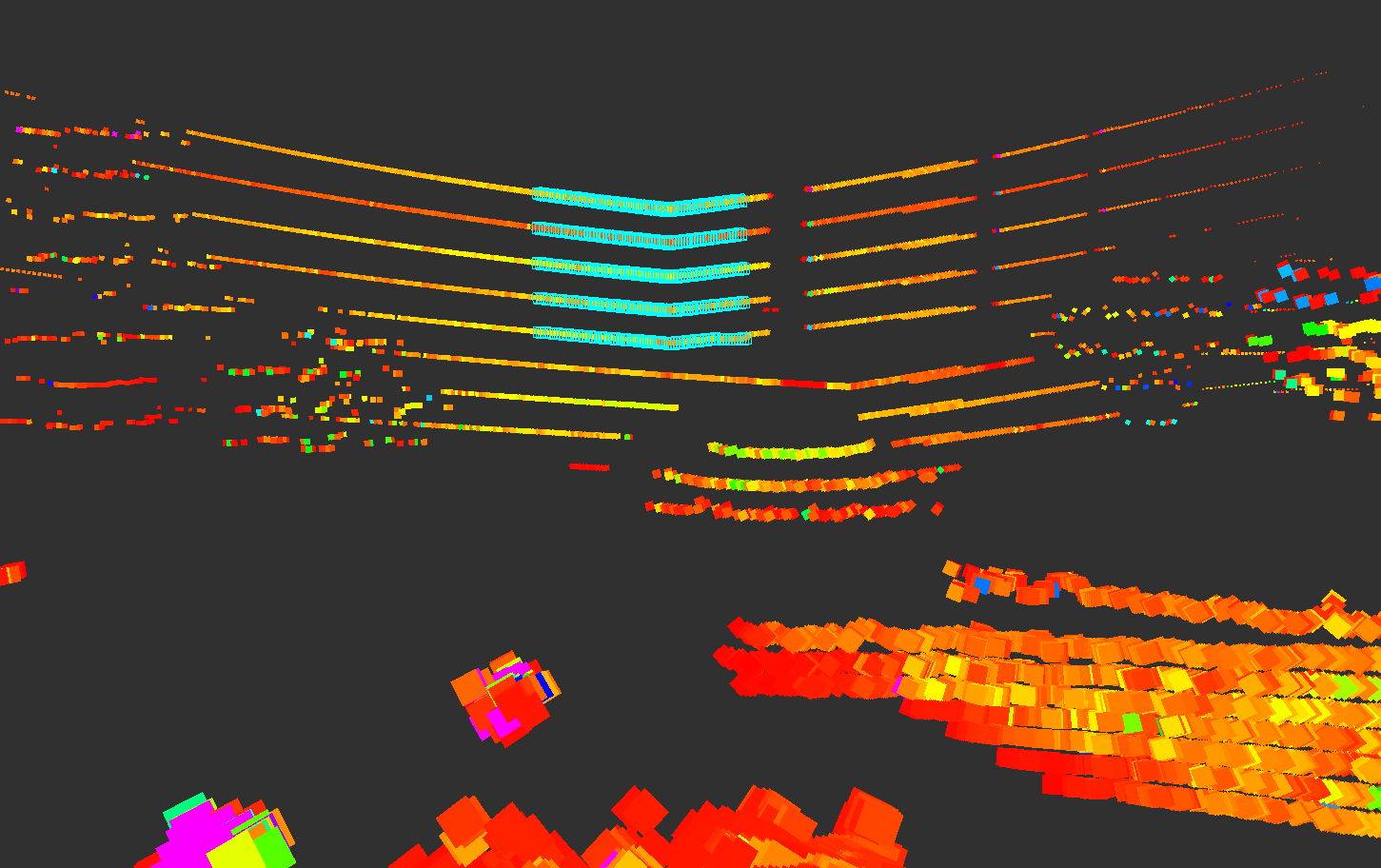} \\[+23pt]
        \small{(a) Segmented building corner (in blue)}
    \end{minipage}
    \begin{minipage}{0.55\textwidth}
        \centering
        \setlength\figureheight{.6\textwidth}
        \setlength\figurewidth{\textwidth}
        \small{
\begin{tikzpicture}

\definecolor{color0}{rgb}{0.929411764705882,0.541176470588235,0}

\begin{axis}[
axis background/.style={fill=white!89.80392156862746!black},
axis line style={white},
height=\figureheight,
tick align=outside,
tick pos=both,
width=\figurewidth,
x grid style={white},
xlabel={Total Horizontal Wall Length (m)},
xmajorgrids,
xmin=-0.2, xmax=3.2,
xtick style={color=white!33.33333333333333!black},
y grid style={white},
ylabel={Segmented PC Size (\#points)},
ymajorgrids,
ymin=30, ymax=320,
ytick style={color=white!33.33333333333333!black}
]
\addplot [semithick, color0, mark=*, mark size=3, mark options={solid}]
table {%
2.6 278
2.2 229
1.96 202
1.55 166
1.36 143
1.21 128
0.9 97
0.46 58
};
\end{axis}

\end{tikzpicture}} \\
        \small{(b) Size of the point cloud for different wall lengths}
    \end{minipage}
    \caption{Size of 3D lidar data samples of a building corner (point cloud data). The size is shown in relation to the total horizontal wall side included in the sample, obtaining a linear relation. Data described in \cite{qingqing2019urbanlocalization}.}
    \label{fig:lidarcorner}
\end{figure*}

We have utilized 3D lidar data from a 32-channel Velodyne laser scanner, which has been previously utilized for localization in \cite{qingqing2019urbanlocalization}, where the platform utilized for gathering the data is described in detail. We have focused on analyzing how the data can be characterized based on the distance to the object that is being utilized for the data sample. In Figure~\ref{fig:lidarcorner} we show a corner of a building. Because the corner is only seen as such when the robot is able to see both walls, and the distance from any point in the road where it is visible is almost constant, the main factor to take into account is the area of the wall that is being considered for the data stamp. We have found that the point density is almost constant, and therefore Figure~\ref{fig:lidarcorner} shows a linear relationship between the horizontal wall distance in the data sample and the number of points in the cloud. Here, we are assuming that the number of channels in the lidar is known, as it can be easily extracted from the point cloud.

\begin{figure*}
    \centering
    \begin{minipage}{0.49\textwidth}
        \centering
        \includegraphics[width=.9\textwidth]{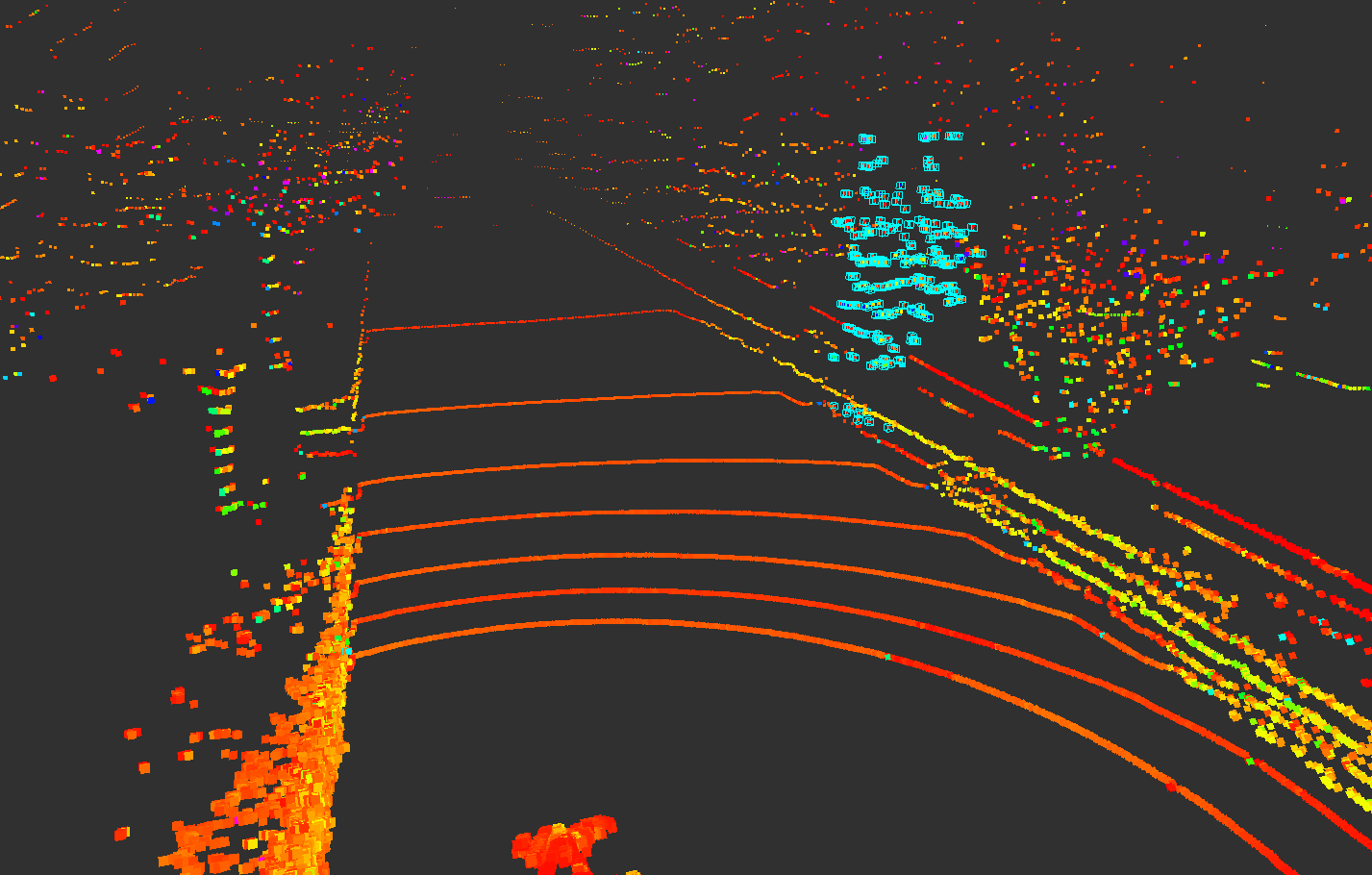} \\
        \small{(a) Segmented tree at 11.8~m (in blue)}
    \end{minipage}
    \begin{minipage}{0.49\textwidth}
        \centering
        \includegraphics[width=.9\textwidth]{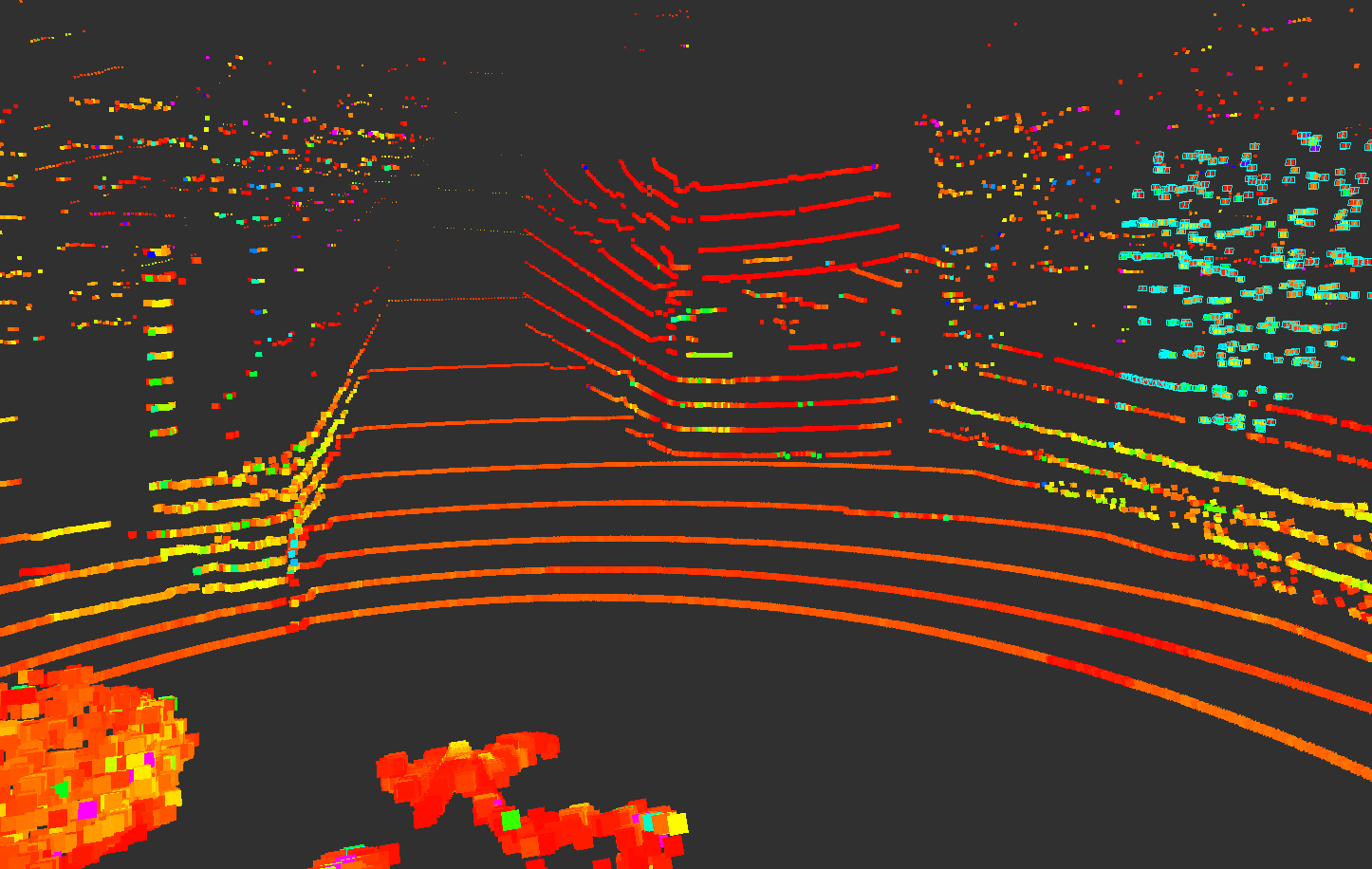} \\
        \small{(b)  Segmented tree at 8~m (in blue)}
    \end{minipage} \\
    $ $ \\[+23pt]
    \begin{minipage}{0.98\textwidth}
        \centering
        \setlength\figureheight{0.42\textwidth}
        \setlength\figurewidth{.8\textwidth}
        \small{
\begin{tikzpicture}

\definecolor{color0}{rgb}{0.929411764705882,0.541176470588235,0}

\begin{axis}[
axis background/.style={fill=white!89.80392156862746!black},
axis line style={white},
height=\figureheight,
tick align=outside,
tick pos=both,
width=\figurewidth,
x grid style={white},
xlabel={Distance to Object (m)},
xmajorgrids,
xmin=6, xmax=18,
xtick style={color=white!33.33333333333333!black},
y grid style={white},
ylabel={Segmented PC Size (\#points)},
ymajorgrids,
ymin=100, ymax=420,
ytick style={color=white!33.33333333333333!black}
]
\addplot [semithick, color0, mark=*, mark size=3, mark options={solid}]
table {%
16.1 134
15.2 165
11.8 241
10.2 326
8 395
};
\end{axis}

\end{tikzpicture}} \\
        \small{(c) Size of the segmented tree point cloud seen at different distances}
    \end{minipage}
    \caption{Size of 3D lidar data samples for a tree in a road (point cloud data). The number of points increases linearly as the sensor is closer to the tree. Data described in \cite{qingqing2019urbanlocalization}.}
    \label{fig:lidartree}
\end{figure*}

\begin{figure*}
    \centering
    \begin{minipage}{0.33\textwidth}
        \centering
        \includegraphics[width=.95\textwidth]{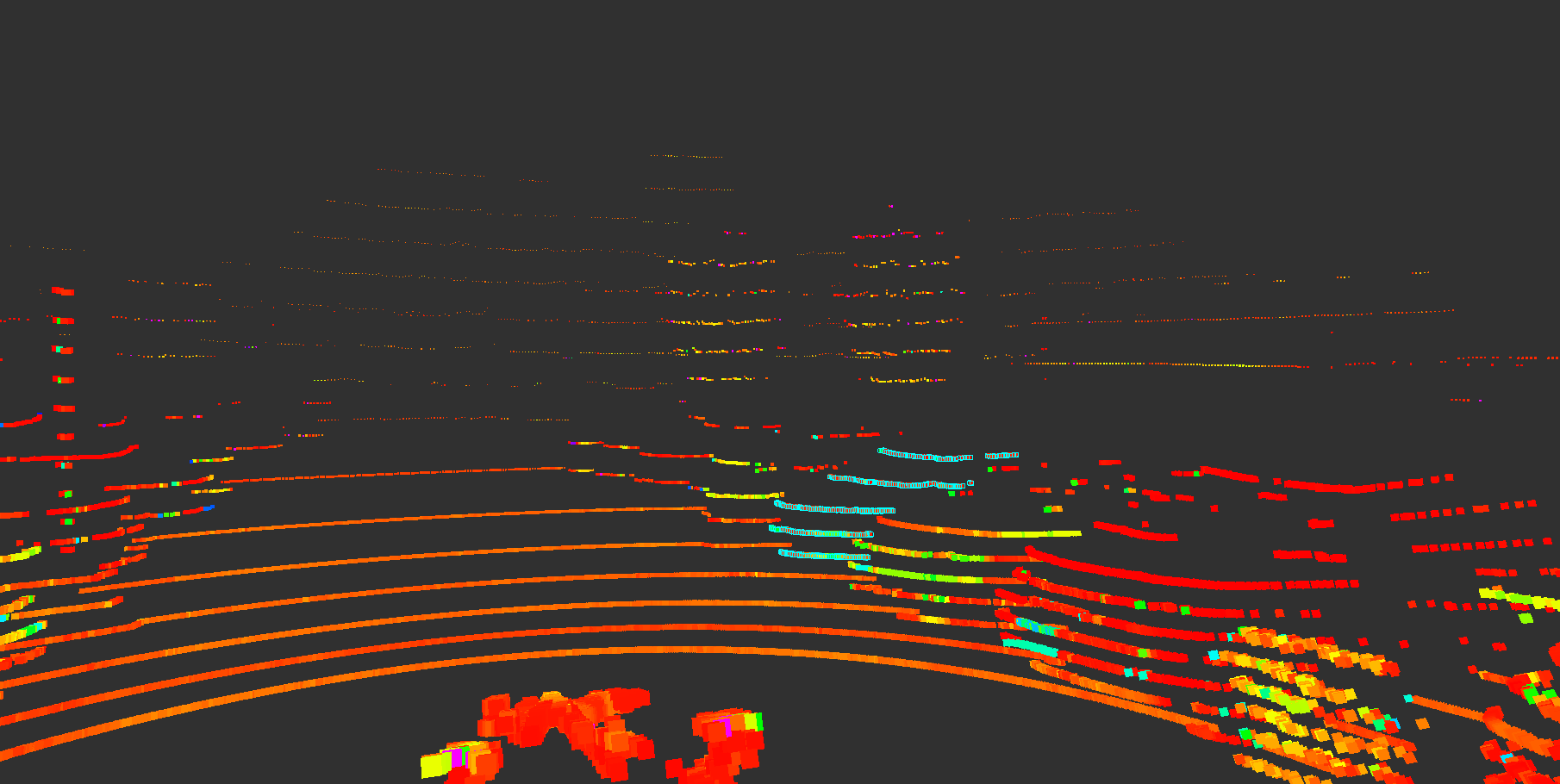} \\
        \small{(a) Car at 8~m (in blue)}
    \end{minipage}
    \begin{minipage}{0.33\textwidth}
        \centering
        \includegraphics[width=.95\textwidth]{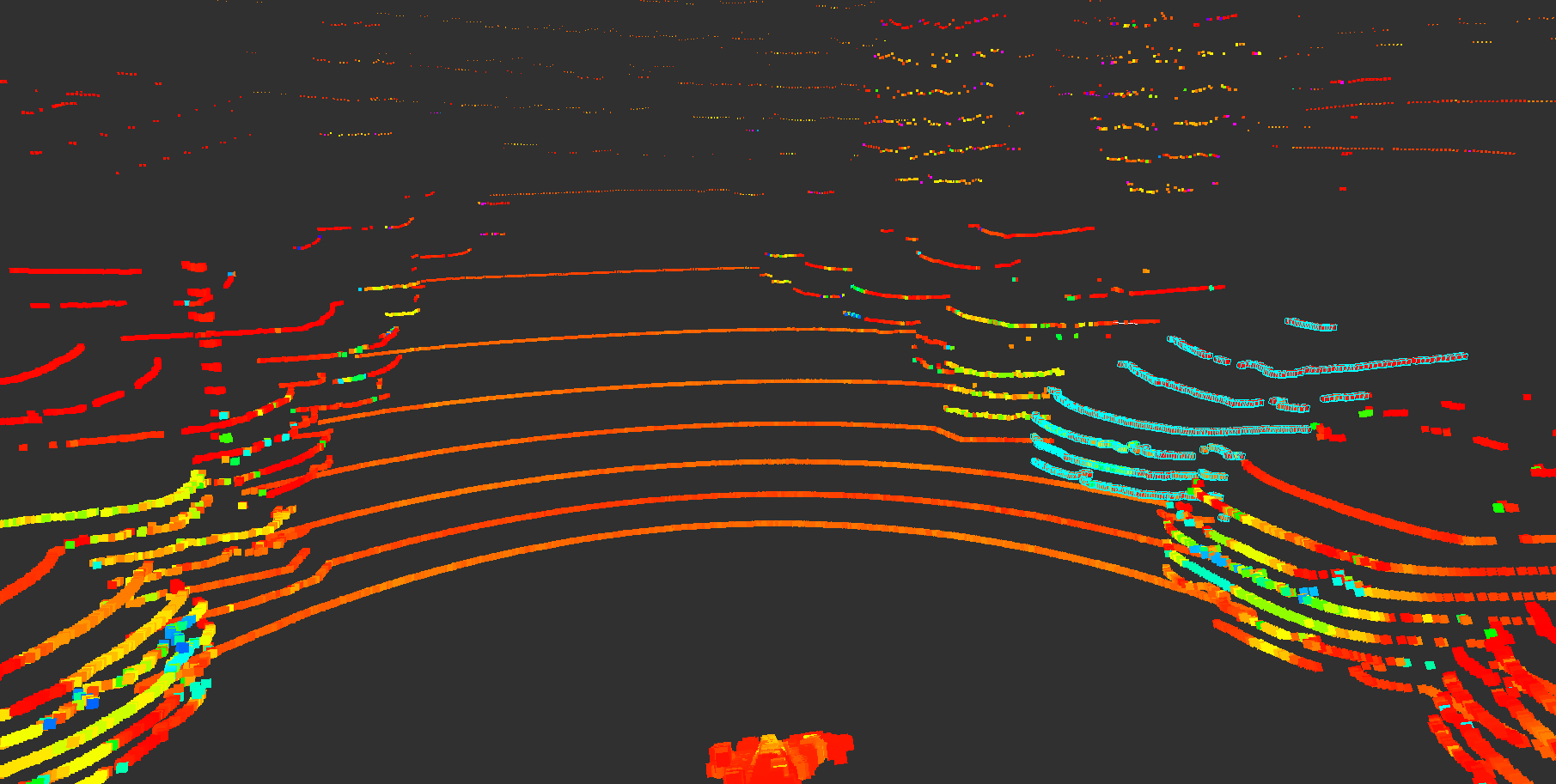} \\
         \small{(b)  Car at 5.5~m (in blue)}
    \end{minipage}
    \begin{minipage}{0.32\textwidth}
        \centering
        \includegraphics[width=.95\textwidth]{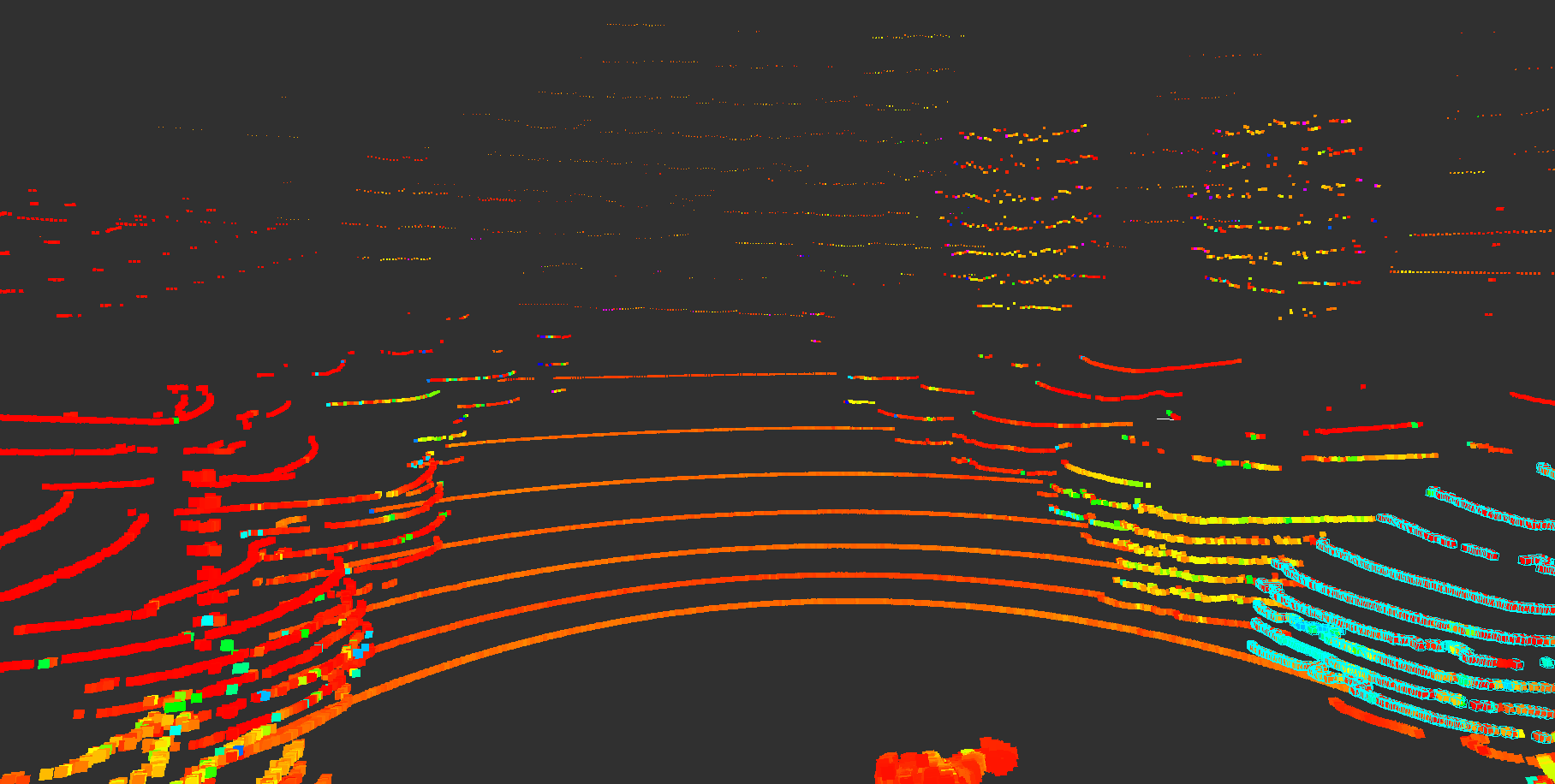} \\
         \small{(c)  Car at 4~m (in blue)}
    \end{minipage} \\
    $ $ \\[+23pt]
    \begin{minipage}{0.98\textwidth}
        \centering
        \setlength\figureheight{0.5\textwidth}
        \setlength\figurewidth{.9\textwidth}
        \small{
\begin{tikzpicture}

\definecolor{color1}{rgb}{1,0.196078431372549,0}
\definecolor{color0}{rgb}{0.929411764705882,0.541176470588235,0}

\begin{axis}[
axis background/.style={fill=white!89.80392156862746!black},
axis line style={white},
height=\figureheight,
legend cell align={left},
legend style={fill opacity=0.8, draw opacity=1, text opacity=1, draw=white!80.0!black, fill=white!89.80392156862746!black},
tick align=outside,
tick pos=both,
width=\figurewidth,
x grid style={white},
xlabel={Distance to Object (m)},
xmajorgrids,
xmin=3.1, xmax=10.5,
xtick style={color=gray!66.66666666666666!black},
y grid style={white},
ylabel={Segmented PC Size (\#points)},
ymajorgrids,
ymin=50, ymax=1150,
ytick style={color=gray!66.66666666666666!black}
]
\addplot [semithick, color0, mark=*, mark size=3, mark options={solid}]
table {%
9.8 150
8 273
5.5 578
4 1117
};
\addlegendentry{16 channels}
\addplot [semithick, color1, mark=*, mark size=3, mark options={solid}]
table {%
9.8 76
8 114
5.5 294
4 457
};
\addlegendentry{8 channels}
\end{axis}

\end{tikzpicture}} \\
        \small{(d) Size of the segmented tree point cloud seen at different distances}
    \end{minipage}
    \caption{Size of 3D lidar data samples for a car in a roadside parking (point cloud data). The number of points does not follow a linear ratio with the distance because the area of the car visible to the sensor increases with the distance, together with the density of the point cloud. We show the number of points in the data samples for a 16-channel and a (simulated) 8-channel lidar, where the number of points depends on the number of channels that are projected onto the car surface. Data described in \cite{qingqing2019urbanlocalization}.}
    \label{fig:lidarcar}
\end{figure*}

Two more types of features are shown in Figures~\ref{fig:lidartree} and~\ref{fig:lidarcar}. First, we analyze the extracted point cloud corresponding to a tree near the road in Figure~\ref{fig:lidartree}. In this case, the surface of the tree that it is visible to the sensor remains almost constant because there are no other objects in the vicinity and due to the shape of the tree being similar to that of a solid of revolution. As the graphic shows, the relationship between the number of points in the segmented tree and the distance to the tree is mostly linear, as the visible area remains similar from different points of view. Finally, Figure~\ref{fig:lidarcar} shows the extracted point cloud from a car in a parking lot next to the road. In this case, both the distance to the object and the visible are important factors because there are other objects nearby. Therefore, different parts of the car are visible from different points of view, and the relationship between the segmented point cloud size and the distance to the object is no longer linear. In this case, moreover, the number of channels that have a projection over the car varies with the distance. We have extracted only eight of the channels from the lidar data to show that ratio between a 16-channel and an 8-channel point cloud size is not constant.

Taking into account the above data, in order to design the data characterization function different types of objects or features and their characteristics must be considered. From these, we can then define a function that defines the data density based on the distance to an object (as in the tree), the area of the feature being encoded in the data sample (as in the case of the building corner), or both the distance and the visible area (as with the car). These are just some examples and by no means do they provide an exhaustive classification. However, in this paper, we have focused on showing different possibilities to define the data characterization scheme and justified them with different real-world cases.

\subsection{Initial Implementation}

An initial implementation with Ethereum has been made, where the data samples are submitted as data payloads in individual transactions between devices. In this setting, smart contracts are not yet utilized to implement the data ranking. Instead, the data is submitted as a payload in standard Ethereum transactions together with 1 Ether. To do this, a private Ethereum network has been deployed with PoW difficulty set to 0 bits. Therefore, the mined blocks and the Ethereum generated is only limited by the time between blocks specified in the genesis file. Table~\ref{tab:ethereum} shows the gas necessary to submit a transaction for inclusion in a blockchain block depending on the size of the data payload. This relationship must be taken into account when deciding the amount of ether or gas given to each robot based on the amount of data that they are going to share with their collaborating peers. Then, the amount of data in each data sample can be predefined based on the total amount of data shared, and checked by the nodes receiving the data. This helps ensuring fair use of the bandwidth in the peer-to-peer network.

\begin{table*}
    \centering
    \caption{Gas needed in order to attach a given ammount of data to a Ethereum transaction. All transactions involved 1 ether and a data payload.}
    \small
    \begin{tabular}{@{}lccccc@{}}
        \toprule
        & Transact. \#1 &  Transact. \#2 &  Transact. \#3 &  Transact. \#4 &  Transact. \#5 \\
        \midrule
        \textbf{Data payload (bytes)}   & 20            & 1080     & 2160               & 4320            & 8640   \\
        \textbf{Gas}                    & 21680         & 57720    & 94440              & 167880          & 314760   \\
        \bottomrule
    \end{tabular}
    \label{tab:ethereum}
\end{table*}

\section{Discussion}

In general terms, the approach proposed in this paper requires more maturity of some blockchain technologies that are not widely in use yet. Some of these can be seen within the roadmap towards Ethereum 2.0. The collaboration process described in the previous section can be summarized with the following steps:
\begin{enumerate}
    \item Genesis of the blockchain. A decision is taken regarding whether it is supported by fixed assets, such as infrastructure, or automatically destroyed when the number of collaborating robots reaches a minimum threshold.
    \item A new robot is able to join the network by providing a PoW solution, in order to avoid Sybil attacks and ensure that all robots have a minimum of available computational resources. Upon joining, the bandwidth of the connection between the robots and the peers that it is connected to is put to test. This can be done periodically in order to have an estimation of the peer-to-peer network bandwidth, if other means of calculating its capacity are not available.
    \item Periodic submission of partial or full PoW solutions. This is utilized to have an online estimation of the available computational resources at the different robots. Together with the PoW solution, a series of data stamp is submitted. Each data stamp must represent the type and density of data that the robot is able to share with the rest of the network. An estimation of the maximum data throughput that it can stream for each of the types must be included as well. If the robot is located in a location where previous data stamps exist, and it is able to capture comparable data, then additional data stamps are submitted as well. These additional stamps must be accompanied by a comparison result and corresponding details, which must be verifiable by robots with enough computing power.
    \item Together with the PoW and data stamps, robots inform their peers about the type of data and a range of resolutions that would benefit their autonomous operation.
    \item If data stamps which can be comparable to previous entries in the blockchain are submitted, then all robots with enough processing capabilities must perform the comparison. An PoS approach is utilized in order to validate a comparison, where the stake is calculated based on the number of positively confirmed data stamps submitted by each individual robot.
    \item Upon receiving all partial PoW and the corresponding data stamps, a smart contract is executed to perform the online estimation of available processing power.
    \item Utilizing the processing capabilities at the receiving robots and the peer-to-peer network bandwidth as constraints, an optimization problem is solved in order to decide the usage of the network. The function to be optimized is a weighted sum of the 
    \item Robots receiving data must confirm that its properties are equivalent to those of the submitted data stamp. In the event of a mismatch or the inability of the receiving robot to verify that the data stamp is part of the received data, a negative receipt is issued which affects the evaluation of data quality and trustability together with the stamp-to-stamp matching process.
\end{enumerate}

\begin{figure*}
    \centering
    \includegraphics[width=\textwidth]{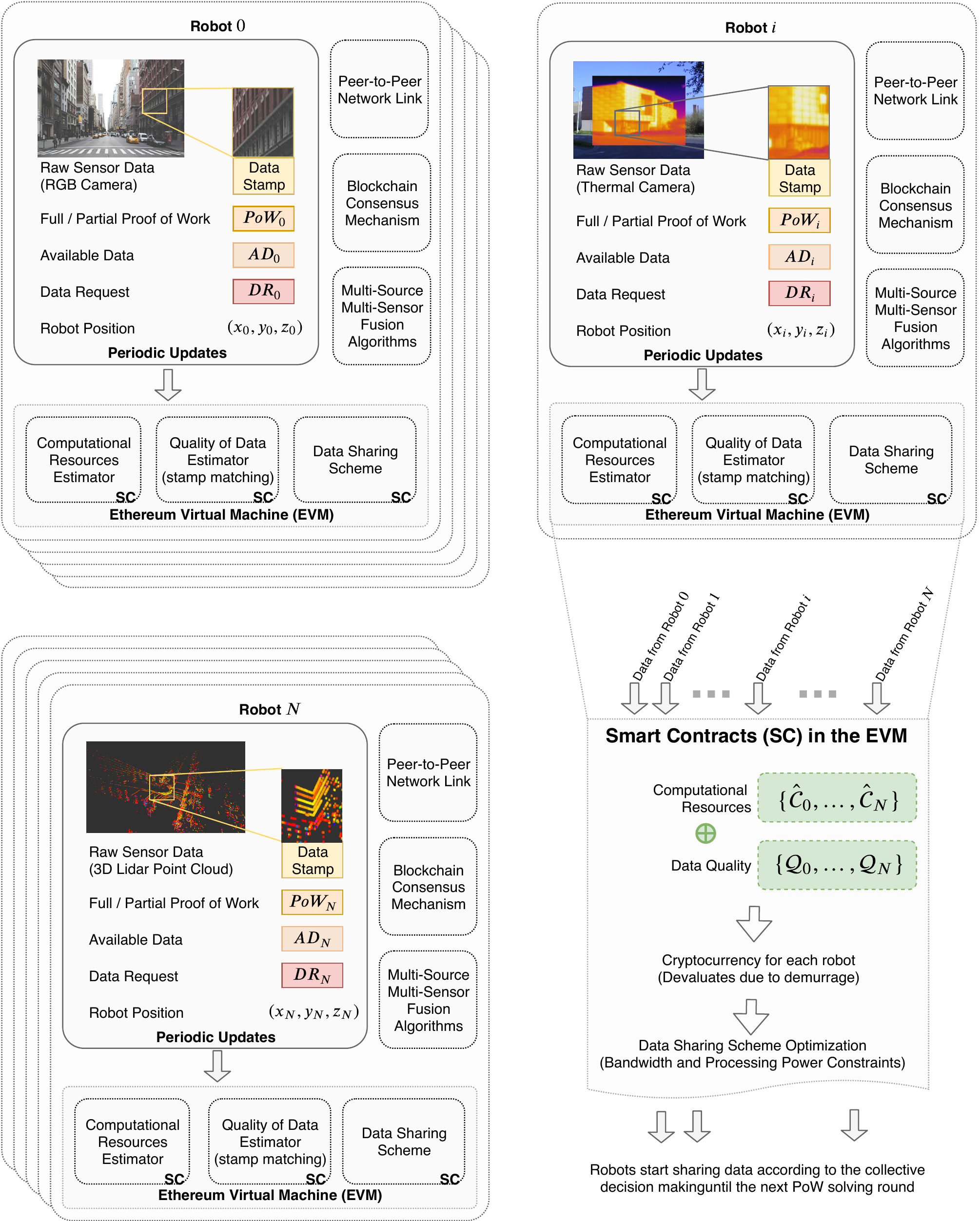}
    \caption{Illustration of the inputs utilized in order to estimate computational resources and data quality. The data stamps are taken from specific features from the environment that can be easily recognized by other robots, such as the structure of windows or the corners in a building.}
    \label{fig:collaborating_chain}
\end{figure*}

An illustration of some of the steps summarized above is provided in Figure \ref{fig:collaborating_chain}, where different robots have different computational capabilities and share data stamps, partial or full PoW solutions, data requests and information about available data in order to collaborative decide on the best usage of the peer-to-peer network and the data that will be shared. After this, robots start collaborating, utilizing the newly coined cryptocurrency to make data exchange transactions until the next PoW puzzle has to be solved. Therefore, the behaviour of the collaborating robots in terms of communication and data sharing is static between two consecutive PoW solving rounds.

\subsection{Challenges}

In order to arrive to a successful implementation of the strategies presented in this paper, numerous challenges need to be overcome. Nonetheless, the main objective of this paper is to define a basic set of design strategies and architectures that have potential to enable secure and efficient ad hoc robotic collaboration in the near future.

Some of the main challenges of the proposed, which will be studied further in future work, are the following: (i) the management of blockchain lifecycles, or the management of trust where a predefined set of assets is in charge of maintaining the blockchain; (ii) the prevention of broad-scale attacks where a large group of attackers submit counterfeit data to the network and continuously provide new stamps and confirmations of such data; and (iii) how to ensure that robots do not join the network only to obtain data but do not show the real quality of their data, for instance by downsampling point clouds or inducing blur in images. This last point presents particular complexity because it is virtually impossible for robots to evaluate whether their peers do have more sensing capabilities or not, and limiting the amount of data that robots receive based on the quality of data that they provide would have a significant negative impact in resource-constrained robots with limited sensor suites such as drones or small delivery robots.

\subsection{Opportunities}

We see the main opportunities of the proposed system as part of smart cities, where the blockchain can be supported by deployed infrastructure, as well as industrial environments where there exists trust between parties but different actors deploy robots in the same environment. In a smart city, a public blockchain for data sharing could boost the deployment of autonomous robots from both private and public entities. In addition, the role of the infrastructure should be considered not only as a platform to manage the blockchain lifecycle but also as a static data source and validating platform, where traffic cameras and other sensors that already exist can be integrated. With this sensor data, connected infrastructure can validate data stamps from the robots, but also provide data to interested parties.
\section{Conclusion}

We have introduced strategies for managing collaboration in ad hoc heterogeneous robotic swarms with blockchain technology. We have analyzed consensus protocols and scalability issues in various blockchains architectures, and proposed different ways of adapting these methods and architectures to manage two parameters in robotic collaboration: the evaluation of the quality of data, and the estimation of computational resources at different robots. With the proposed approach, these parameters can be estimated without utilizing explicit information about the sensors and hardware onboard the robots. In particular, we propose the utilization of periodic PoW to estimate the relative available computational resources at different robots. In addition, when robots share data they must include a certain sample through a transaction in the blockchain, such that it can be utilized by other robots operating in the environment simultaneously or in the future in order to confirm that the data is both valid and of good quality. We have argued that this approach has potential to be utilized in smart cities, where connected infrastructure is in charge of maintaining the blockchain even if no robots are connected.

Some of the concepts presented in this paper are in early development state. Consequently, a full and efficient implementation of the proposed strategies is not viable with the current state-of-the-art. Furthermore, multiple challenges remain in the definition of generic yet flexible and adaptable methods for evaluating the data quality and increasing the efficacy of the ranking scheme. In future work, we will study these aspects more in depth, as well as provide partial implementation and testing with specific application scenarios.


\bibliographystyle{unsrt}
\bibliography{bibliography}

\begin{thebibliography}{10}

\bibitem{van2016healthcare}
Aimee Van~Wynsberghe.
\newblock {\em Healthcare robots: Ethics, design and implementation}.
\newblock Routledge, 2016.

\bibitem{bojarski2016end}
Mariusz Bojarski, Davide Del~Testa, Daniel Dworakowski, Bernhard Firner, Beat
  Flepp, Prasoon Goyal, Lawrence~D Jackel, Mathew Monfort, Urs Muller, Jiakai
  Zhang, et~al.
\newblock End to end learning for self-driving cars.
\newblock {\em arXiv preprint arXiv:1604.07316}, 2016.

\bibitem{russmann2015industry}
Michael R{\"u}{\ss}mann, Markus Lorenz, Philipp Gerbert, Manuela Waldner, Jan
  Justus, Pascal Engel, and Michael Harnisch.
\newblock Industry 4.0: The future of productivity and growth in manufacturing
  industries.
\newblock {\em Boston Consulting Group}, 9(1):54--89, 2015.

\bibitem{yaghoubi2013autonomous}
Sajjad Yaghoubi, Negar~Ali Akbarzadeh, Shadi~Sadeghi Bazargani, Sama~Sadeghi
  Bazargani, Marjan Bamizan, and Maryam~Irani Asl.
\newblock Autonomous robots for agricultural tasks and farm assignment and
  future trends in agro robots.
\newblock {\em International Journal of Mechanical and Mechatronics
  Engineering}, 13(3):1--6, 2013.

\bibitem{kaneko2008humanoid}
Kenji Kaneko, Kensuke Harada, Fumio Kanehiro, Go~Miyamori, and Kazuhiko Akachi.
\newblock Humanoid robot hrp-3.
\newblock In {\em 2008 IEEE/RSJ International Conference on Intelligent Robots
  and Systems}, pages 2471--2478. IEEE, 2008.

\bibitem{alami1998cooperation}
R.~{Alami}, S.~{Fleury}, M.~{Herrb}, F.~{Ingrand}, and F.~{Robert}.
\newblock Multi-robot cooperation in the martha project.
\newblock {\em IEEE Robotics Automation Magazine}, 5(1):36--47, March 1998.

\bibitem{dadgar2016pso}
Masoud Dadgar, Shahram Jafari, and Ali Hamzeh.
\newblock A pso-based multi-robot cooperation method for target searching in
  unknown environments.
\newblock {\em Neurocomputing}, 177:62--74, 2016.

\bibitem{gross2006autonomous}
Roderich Gro{\ss}, Michael Bonani, Francesco Mondada, and Marco Dorigo.
\newblock Autonomous self-assembly in swarm-bots.
\newblock {\em IEEE transactions on robotics}, 22(6):1115--1130, 2006.

\bibitem{rubenstein2012kilobot}
Michael Rubenstein, Christian Ahler, and Radhika Nagpal.
\newblock Kilobot: A low cost scalable robot system for collective behaviors.
\newblock In {\em 2012 IEEE International Conference on Robotics and
  Automation}, pages 3293--3298. IEEE, 2012.

\bibitem{shi2010task}
Zhiguo Shi, Junming Wei, Xujian Wei, Kun Tan, and Zhiliang Wang.
\newblock The task allocation model based on reputation for the heterogeneous
  multi-robot collaboration system.
\newblock In {\em 2010 8th World Congress on Intelligent Control and
  Automation}, pages 6642--6647. IEEE, 2010.

\bibitem{qin2016recent}
Jiahu Qin, Qichao Ma, Yang Shi, and Long Wang.
\newblock Recent advances in consensus of multi-agent systems: A brief survey.
\newblock {\em IEEE Transactions on Industrial Electronics}, 64(6):4972--4983,
  2016.

\bibitem{sankar2017survey}
Lakshmi~Siva Sankar, M~Sindhu, and M~Sethumadhavan.
\newblock Survey of consensus protocols on blockchain applications.
\newblock In {\em 2017 4th International Conference on Advanced Computing and
  Communication Systems (ICACCS)}, pages 1--5. IEEE, 2017.

\bibitem{wang2019survey}
Wenbo Wang, Dinh~Thai Hoang, Peizhao Hu, Zehui Xiong, Dusit Niyato, Ping Wang,
  Yonggang Wen, and Dong~In Kim.
\newblock A survey on consensus mechanisms and mining strategy management in
  blockchain networks.
\newblock {\em IEEE Access}, 7:22328--22370, 2019.

\bibitem{ferrer2018blockchain}
Eduardo Castell{\'o}~Ferrer.
\newblock The blockchain: a new framework for robotic swarm systems.
\newblock In {\em Proceedings of the Future Technologies Conference}, pages
  1037--1058. Springer, 2018.

\bibitem{yuan2018blockchain}
Yong Yuan and Fei-Yue Wang.
\newblock Blockchain and cryptocurrencies: Model, techniques, and applications.
\newblock {\em IEEE Transactions on Systems, Man, and Cybernetics: Systems},
  48(9):1421--1428, 2018.

\bibitem{strobel2018managing}
Volker Strobel, Eduardo Castell{\'o}~Ferrer, and Marco Dorigo.
\newblock Managing byzantine robots via blockchain technology in a swarm
  robotics collective decision making scenario.
\newblock In {\em AAMAS}, pages 541--549, 2018.

\bibitem{nguyen2019blockchain}
Trung~T Nguyen, Amartya Hatua, and Andrew~H Sung.
\newblock Blockchain approach to solve collective decision making problems for
  swarm robotics.
\newblock In {\em International Congress on Blockchain and Applications}, pages
  118--125. Springer, 2019.

\bibitem{castello2019merkle}
Eduardo Castell{\'o}~Ferrer, Thomas Hardjono, and Alex Pentland.
\newblock Secure and secret cooperation of robotic swarms by using merkle
  trees.
\newblock {\em CoRR}, abs/1904.09266, 2019.

\bibitem{queralta2019collaborative}
Jorge Peña~Queralta, Tuan~Nguyen Gia, Hannu Tenhunen, and Tomi Westerlund.
\newblock Collaborative mapping with ioe-based heterogeneous vehicles for
  enhanced situational awareness.
\newblock In {\em 2019 IEEE Sensors Applications Symposium (SAS)}, pages 1--6.
  IEEE, 2019.

\bibitem{cardona2019robot}
Gustavo~A Cardona and Juan~M Calderon.
\newblock Robot swarm navigation and victim detection using rendezvous
  consensus in search and rescue operations.
\newblock {\em Applied Sciences}, 9(8):1702, 2019.

\bibitem{zheng2017overview}
Zibin Zheng, Shaoan Xie, Hongning Dai, Xiangping Chen, and Huaimin Wang.
\newblock An overview of blockchain technology: Architecture, consensus, and
  future trends.
\newblock In {\em 2017 IEEE International Congress on Big Data (BigData
  Congress)}, pages 557--564. IEEE, 2017.

\bibitem{luu2016secure}
Loi Luu, Viswesh Narayanan, Chaodong Zheng, Kunal Baweja, Seth Gilbert, and
  Prateek Saxena.
\newblock A secure sharding protocol for open blockchains.
\newblock In {\em Proceedings of the 2016 ACM SIGSAC Conference on Computer and
  Communications Security}, pages 17--30. ACM, 2016.

\bibitem{kokoris2018omniledger}
Eleftherios Kokoris-Kogias, Philipp Jovanovic, Linus Gasser, Nicolas Gailly,
  Ewa Syta, and Bryan Ford.
\newblock Omniledger: A secure, scale-out, decentralized ledger via sharding.
\newblock In {\em 2018 IEEE Symposium on Security and Privacy (SP)}, pages
  583--598. IEEE, 2018.

\bibitem{wood2014ethereum}
Gavin Wood et~al.
\newblock Ethereum: A secure decentralised generalised transaction ledger.
\newblock {\em Ethereum project yellow paper}, 151(2014):1--32, 2014.

\bibitem{li2017survey}
Xiaoqi Li, Peng Jiang, Ting Chen, Xiapu Luo, and Qiaoyan Wen.
\newblock A survey on the security of blockchain systems.
\newblock {\em Future Generation Computer Systems}, 2017.

\bibitem{nakamoto2008bitcoin}
Satoshi Nakamoto et~al.
\newblock {\em Bitcoin: A peer-to-peer electronic cash system}.
\newblock Working Paper, 2008.

\bibitem{dwork1992pricing}
Cynthia Dwork and Moni Naor.
\newblock Pricing via processing or combatting junk mail.
\newblock In {\em Annual International Cryptology Conference}, pages 139--147.
  Springer, 1992.

\bibitem{eyal2015miner}
Ittay Eyal.
\newblock The miner's dilemma.
\newblock In {\em 2015 IEEE Symposium on Security and Privacy}, pages 89--103.
  IEEE, 2015.

\bibitem{zhu2018survey}
Saide Zhu, Wei Li, Hong Li, Chunqiang Hu, and Zhipeng Cai.
\newblock A survey: Reward distribution mechanisms and withholding attacks in
  bitcoin pool mining.
\newblock {\em Mathematical Foundations of Computing}, 1(4):393--414, 2018.

\bibitem{schrijvers2016incentive}
Okke Schrijvers, Joseph Bonneau, Dan Boneh, and Tim Roughgarden.
\newblock Incentive compatibility of bitcoin mining pool reward functions.
\newblock In {\em International Conference on Financial Cryptography and Data
  Security}, pages 477--498. Springer, 2016.

\bibitem{miller2015nonoutsourceable}
Andrew Miller, Ahmed Kosba, Jonathan Katz, and Elaine Shi.
\newblock Nonoutsourceable scratch-off puzzles to discourage bitcoin mining
  coalitions.
\newblock In {\em Proceedings of the 22nd ACM SIGSAC Conference on Computer and
  Communications Security}, pages 680--691. ACM, 2015.

\bibitem{barber2012bitter}
Simon Barber, Xavier Boyen, Elaine Shi, and Ersin Uzun.
\newblock Bitter to better—how to make bitcoin a better currency.
\newblock In {\em International Conference on Financial Cryptography and Data
  Security}, pages 399--414. Springer, 2012.

\bibitem{popov2016probabilistic}
Serguei Popov.
\newblock A probabilistic analysis of the nxt forging algorithm.
\newblock {\em Ledger}, 1:69--83, 2016.

\bibitem{nxt2018whitepaper}
Nxt Wiki.
\newblock {\em Whitepaper: Nxt}.
\newblock Nxtwiki. org [online] https://nxtwiki. org, 2018.

\bibitem{bentov2016cryptocurrencies}
Iddo Bentov, Ariel Gabizon, and Alex Mizrahi.
\newblock Cryptocurrencies without proof of work.
\newblock In {\em International Conference on Financial Cryptography and Data
  Security}, pages 142--157. Springer, 2016.

\bibitem{buterin2019incentives}
Vitalik Buterin, Daniel Reijsbergen, Stefanos Leonardos, and Georgios
  Piliouras.
\newblock Incentives in ethereum's hybrid casper protocol.
\newblock {\em arXiv preprint arXiv:1903.04205}, 2019.

\bibitem{hildenbrandt2018kevm}
Everett Hildenbrandt, Manasvi Saxena, Nishant Rodrigues, Xiaoran Zhu, Philip
  Daian, Dwight Guth, Brandon Moore, Daejun Park, Yi~Zhang, Andrei Stefanescu,
  et~al.
\newblock Kevm: A complete formal semantics of the ethereum virtual machine.
\newblock In {\em 2018 IEEE 31st Computer Security Foundations Symposium
  (CSF)}, pages 204--217. IEEE, 2018.

\bibitem{solidity}
{Ethereum Revision 7709ece9}.
\newblock {\em Solidity Documentation}.
\newblock Solidity Read The Docs [online]
  https://solidity.readthedocs.io/en/v0.5.12/. Accessed October 2019.,
  2016-2019.

\bibitem{dannen2017introducing}
Chris Dannen.
\newblock {\em Introducing Ethereum and Solidity}.
\newblock Springer, 2017.

\bibitem{vujivcic2018blockchain}
Dejan Vuji{\v{c}}i{\'c}, Dijana Jagodi{\'c}, and Sini{\v{s}}a Ran{d}i{\'c}.
\newblock Blockchain technology, bitcoin, and ethereum: A brief overview.
\newblock In {\em 2018 17th International Symposium INFOTEH-JAHORINA
  (INFOTEH)}, pages 1--6. IEEE, 2018.

\bibitem{vukolic2015quest}
Marko Vukoli{\'c}.
\newblock The quest for scalable blockchain fabric: Proof-of-work vs. bft
  replication.
\newblock In {\em International workshop on open problems in network security},
  pages 112--125. Springer, 2015.

\bibitem{ethereum20specification}
Serenity Ethereum~Foundation et~al.
\newblock {\em Ethereum 2.0 Specifications}.
\newblock Accessed October 2019 [online]
  https://github.com/ethereum/eth2.0-specs, 2018.

\bibitem{ethereum2017goals}
Ben Edington.
\newblock {\em Exploring Ethereum 2.0 Design Goals}.
\newblock Consensys. Accessed October 2019 [online]
  https://media.consensys.net/exploring-the-ethereum-2-0-design-goals-fd2d901b4c01,
  January 2017.

\bibitem{ethereum20serenity}
EthHub.
\newblock {\em Ethereum 2.0 (Serenity) Phases}.
\newblock Accessed October 2019 [online]
  https://docs.ethhub.io/ethereum-roadmap/ethereum-2.0/eth-2.0-phases/, 2018.

\bibitem{state1}
Ben Edington.
\newblock {\em State of Ethereum Protocol \#1}.
\newblock Consensys. Accessed October 2019 [online]
  https://media.consensys.net/state-of-ethereum-protocol-1-d3211dd0f6, August
  2018.

\bibitem{state2}
Ben Edington.
\newblock {\em State of Ethereum Protocol \#2}.
\newblock Consensys. Accessed October 2019 [online]
  https://media.consensys.net/state-of-ethereum-protocol-2-the-beacon-chain-c6b6a9a69129,
  October 2018.

\bibitem{ethereum2018sharding}
{Vitalik Buterin and others}.
\newblock {\em On sharding blockchains}.
\newblock Ethereum Wiki. Sharding FAQ. Accessed October 2019. Version April
  18th, 2019 [online] https://github.com/ethereum/wiki/wiki/Sharding-FAQ,
  {2016-2019}.

\bibitem{arkin1997cooperative}
Ronald~C Arkin and Tucker Balch.
\newblock Cooperative multiagent robotic systems.
\newblock 1997.

\bibitem{quinn2003evolving}
Matt Quinn, Lincoln Smith, Giles Mayley, and Phil Husbands.
\newblock Evolving controllers for a homogeneous system of physical robots:
  Structured cooperation with minimal sensors.
\newblock {\em Philosophical Transactions of the Royal Society of London.
  Series A: Mathematical, Physical and Engineering Sciences},
  361(1811):2321--2343, 2003.

\bibitem{tuci2018cooperative}
Elio Tuci, Muhanad~H Alkilabi, and Otar Akanyeti.
\newblock Cooperative object transport in multi-robot systems: A review of the
  state-of-the-art.
\newblock {\em Frontiers in Robotics and AI}, 5:59, 2018.

\bibitem{zhao2015evolved}
Zeng-Shun Zhao, Xiang Feng, Yan-yan Lin, Fang Wei, Shi-Ku Wang, Tong-Lu Xiao,
  Mao-Yong Cao, and Zeng-Guang Hou.
\newblock Evolved neural network ensemble by multiple heterogeneous swarm
  intelligence.
\newblock {\em Neurocomputing}, 149:29--38, 2015.

\bibitem{akbari2019novel}
Zohreh Akbari and Rainer Unland.
\newblock A novel heterogeneous swarm reinforcement learning method for
  sequential decision making problems.
\newblock {\em Machine Learning and Knowledge Extraction}, 1(2):590--610, 2019.

\bibitem{buterin2017casper}
Vitalik Buterin and Virgil Griffith.
\newblock Casper the friendly finality gadget.
\newblock {\em arXiv preprint arXiv:1710.09437}, 2017.

\bibitem{eyal2018majority}
Ittay Eyal and Emin~G{\"u}n Sirer.
\newblock Majority is not enough: Bitcoin mining is vulnerable.
\newblock {\em Communications of the ACM}, 61(7):95--102, 2018.

\bibitem{rosenfeld2011analysis}
Meni Rosenfeld.
\newblock Analysis of bitcoin pooled mining reward systems.
\newblock {\em arXiv preprint arXiv:1112.4980}, 2011.

\bibitem{krizhevsky2009learning}
Alex Krizhevsky, Geoffrey Hinton, et~al.
\newblock Learning multiple layers of features from tiny images.
\newblock Technical report, Citeseer, 2009.

\bibitem{geiger2013vision}
Andreas Geiger, Philip Lenz, Christoph Stiller, and Raquel Urtasun.
\newblock Vision meets robotics: The kitti dataset.
\newblock {\em The International Journal of Robotics Research},
  32(11):1231--1237, 2013.

\bibitem{qingqing2019urbanlocalization}
Tuan Nguyen Gia Zhuo Zou Tomi~Westerlund Li~Qingqing, Jorge Peña~Queralta.
\newblock Multi sensor fusion for navigation and mapping in autonomous
  vehicles: Accurate localization in urban environments.
\newblock {\em Unmanned Systems}, 2020.
\newblock The 9th IEEE International Conference on Cybernetics and Intelligent
  Systems (CIS) and the 9th IEEE International Conference on Robotics,
  Automation and Mechatronics (RAM).

\end{thebibliography}

\end{document}